# Influence of different factors on survival of patients with colorectal cancer

by
XIE Boda
(18253024)

A thesis submitted in partial fulfillment of the requirements
for the degree of

Bachelor of Science (Honours)
in 2021

at

Hong Kong Baptist University
28 December 2021

# Content

# ABSTRACT

**Rectal cancer refers to the cancer from the dentate line to the junction of rectosigmoid colon, is one of the most common malignant tumors of the digestive tract, and the treatment of rectal cancer is controversial. Therefore, understanding the risk factors and survival factors of rectal cancer is of great significance for the diagnosis of patients. This study sampled patients with rectal cancer from the SEER database. The factors affecting the survival of colorectal cancer patients were analyzed by combining principal component analysis and competitive risk model, and the survival time of patients was analyzed by combining principal component analysis and linear regression. Finally, the data were predicted. The results show that principal component analysis can effectively reduce the number of variables, and the combination of competitive risk model and linear regression model can effectively analyze and predict the data.**

# 1. Introduction

Colorectal cancer, also known as "colorectal cancer", which is the cancer of large intestine epithelial sources, including colon cancer and rectal cancer, pathological type is the most common with adenocarcinoma, very few for scale cancer. Worldwide, colorectal cancer is the most common, followed by colon cancer (sigmoid, cecum, ascending, descending, and transverse colon)

Worldwide, colorectal cancer is the third most common malignancy and the second most common cause of death from malignancies. According to the GLOBOCAN Project of the WHO Cancer Research Centre, the number of new cases of colorectal cancer worldwide in 2018 was about 1.8 million and the number of deaths was about 880,000.
Of large intestine cancer come on and the element such as age, area, sex is concerned.
Age: Colorectal cancer mainly occurs in middle-aged and elderly people over 40 years old;
**Gender**: The incidence of colon cancer in men and women is relatively close, while rectal cancer is more common in men:
**Region**: Colorectal cancer is the main colorectal cancer in China, while colon cancer is the main colorectal cancer in European and American countries.
Survival: in recent years, the incidence and mortality of colorectal cancer have increased significantly, which should be paid enough attention to.

## 1.1 Anatomical site classification

**Rectal cancer**: rectal cancer refers to the cancer from the dentate line to the junction of rectosigmoid colon (rectum about 15-18cm long); Left colon cancer: cancer of the left colon of the abdomen, including left transverse colon cancer, descending colon cancer and sigmoid.
**Colon cancer**: Right colon cancer: cancer of the colon on the right side of the abdomen, including cecum, ascending colon cancer, and right transverse colon cancer.

## 1.2 Pathogenesis

The specific cause of colorectal cancer has not been clear. Colorectal cancer is believed to be the result of a combination of environmental and genetic factors such as eating habits.
**Dietary factors**
The incidence of large intestine cancer and dietary factors are closely related. Low fiber diet, high fat and high protein diet, lack of trace elements and vitamins (including calcium deficiency,

Selenium, molybdenum, antioxidant vitamins ACE and beta-carotene are all risk factors for colorectal cancer.

**Genetic factors**

Genetic factors play an important role in the pathogenesis of colorectal cancer. Of these, familial adenomatous polyposis (FAP) is 100% cancerous. In addition, a family history of colorectal cancer increases the risk of colorectal cancer by four times.

**Chemical carcinogen**

Nitrosamine and its compounds are the most important chemical carcinogens leading to colorectal cancer. Methyl aromatic amine in fried and baked food is also closely related to the occurrence of colorectal cancer. In addition, bile acids and cholesterol can also form a variety of chemical carcinogens under the action of intestinal anaerobic bacteria.

**Digestive tract disease**

Patients with ulcerative colitis, Crohn's disease, colorectal adenoma and rectal polyps also have an increased risk of colorectal cancer later in life.

**Lifestyle**

Tobacco is a definite carcinogen, and smoking is closely related to the occurrence of colorectal adenoma. In addition, obesity, mental stress is also a risk factor for colorectal cancer.

**Parasites**

Schistosomiasis is also thought to be a cause of colorectal cancer, especially in chronic schistosomiasis patients.

## 1.3 Symptoms

The symptom of large intestine cancer is concerned with the development stage of the disease, pathological change place; As the disease progresses, a variety of digestive and systemic symptoms may occur.

Among them, left colon cancer, right colon cancer and rectal cancer clinical symptoms are slightly different. If metastasis occurs, it may cause dysfunction of the transplanted organ. Such as liver function damage after liver metastasis, jaundice, dyspnea after lung metastasis, dizziness, headache and pain of bone metastasis.

## 1.4 Cancer Stage

Generally speaking, there are many staging standards for colorectal cancer, among which the most commonly used is the international TNM staging. According to TNM staging standards, colorectal cancer can be divided into four stages, as follows:

**Stage I** : The infiltration of the cancer reaches the submucosa of the muscularis propria of the intestinal wall without lymph node metastasis or distant metastasis.

**Stage II** : refers to the depth of cancer infiltration that has invaded the muscularis propria, to reach the serous membrane, or the primary focus is located in the colon, rectum without serous membrane layer, cancer invasion and to reach the colon or rectal tissue.
**Stage III** : it refers to stage iii as long as there is lymph node metastasis regardless of the size and invasion depth of the cancer.
**Stage IV** : lesions with distant metastasis.

## 1.5 Survival rate

The survival rate of each stage of colon cancer is different, the survival rate of **stage I** colon cancer is better, for example, the 5-year survival rate is above 90%, the survival rate of **stage II** colon cancer is about 60%-70%, the survival rate of stage III colon cancer is about 50%, if it is **stage IV** colon cancer, that is, the 5-year survival rate of late colon cancer is on the low side, about 10%-20%.
Due to the shallow invasion depth of **stage I** colon cancer and no lymph node metastasis, the survival rate of patients is relatively high and clinical cure can be achieved basically. Stage ii and stage III colon cancer generally need chemotherapy after surgery, and there are many chemotherapy schemes, commonly used capecitabine combined with oxaliplatin chemotherapy scheme, and within 5 years, they need to visit the hospital regularly. If the chemotherapy is effective, the follow-up is very close, and the survival of patients can be appropriately prolonged. But if the tumor progresses during treatment, for example advanced colon cancer, the overall survival of the patient is shortened, which is lower, and for colon cancer itself, which is advanced, the survival of the patient is even lower, because the tumor has metastasized far enough that it can quickly spread throughout the body.

## 1.6 Treatment

The treatment of colorectal cancer should adopt the principle of individual treatment, according to the patient's age, constitution, tumor pathological type, invasion range (stage), choose the appropriate treatment method, in order to maximize the radical cure of tumor, and improve the cure rate.
The scope of colorectal cancer invasion is different, the treatment principles are also different: Carcinoma in situ can be treated under endoscopy, the effect is better, can achieve the effect of radical cure.
**Early colorectal cancer**: Surgical treatment can achieve the purpose of radical cure, partial can also use endoscopic treatment to achieve radical cure.
**In the middle and late stage of colorectal cancer**: The comprehensive treatment is mainly based on surgery, that is, the auxiliary application of chemotherapy and targeted therapy, radiotherapy and other methods after surgery.

For patients with middle and advanced colorectal cancer who cannot undergo surgery, radiotherapy, chemotherapy or targeted therapy can be used according to the disease condition to improve patient survival.

Patients with recurrent or distant metastatic colorectal cancer can be treated with chemotherapy or targeted therapy, and some patients can also be treated with surgery to prolong survival, especially for colon cancer.

Recrudescence or the rectum cancer of transfer, give priority to with put change cure or target cure, do not do an operation commonly.

## 1.7 Our research

Colorectal cancer is the most common malignant tumor in the digestive tract. Different subjective and objective factors will lead to the occurrence and deterioration of colorectal cancer, and there are many treatment methods for colorectal cancer. Therefore, in this paper, samples of patients with colorectal cancer and various factors of patients were extracted from SEER data, and the survival period of patients was predicted by using multiple regression models and competitive risk models using principal component analysis method. After the conclusion is drawn, the multiple regression model and competition model are compared, and whether to use the principal component is compared. We also tested the predicted models to ensure their accuracy. Finally, we predicted the survival time of patients who were still alive in some data to test the correctness of our model again.

# 2. Data preprocessing

## 2.1 Data Processing

We preprocessed the original data, deleted the samples containing "blank" and missing values, and deleted irrelevant variables (patient ID, PRCA 2017, TNM 7/CS V0204+ SCHEMA , etc). Finally, we obtained 119,238 samples of data, including 25 diseases (diseases with less than 100 cases have been grouped into "other diseases"). The data is then further re-encoded. We then recode the data to numerically prepare it for subsequent analysis.

First, we processed the age, changing the "100+" to 100.
We changed the gender to 01 variable, where 1 refers to male, 0 refers to female.

We changed Chemotherapy.recode..yes..no.unk.variable,
SEER.Combined.Mets.at.DX.bone..2010..variable,
SEER.Combined.Mets.at.DX.brain..2010..variable,
SEER.Combined.Mets.at.DX.liver..2010..variable,
SEER.Combined.Mets.at.DX.lung..2010..variable, beam_radiation variable,
Combination_of_beam_with_implants_or_isotopes variable,
Radioactive_implants variable,
Radiation_NOS_method_or_source_not_specified variable,
Radioisotopes variable,
Recommended_unknown_if_administered variable,
Refused variable to 01 variable, where 1 means yes, 0 means no.

We changed Derived.AJCC.M..7th.ed..2010.2015.variable to 01variable, where 1 means "M0", 0 means others.

We distinguish Grade..thru.2017. According to its degree , "grade I" means 1, "grade II means"2, and so on.

We distinguish Derived.AJCC.Stage.Group..7th.ed..2010.2015. Accoding to its degree , where "I" is defined as 1, "II", "IIA", "IIB", "IIC", "IINOS" are defined as 2, and by that analogy others can be defined as 3, 4 and 0.

We distinguish Derived.AJCC.T..7th.ed..2010.2015.variable . According to its degree, where the degree "T1", "T1a", "T1b", "T1NOS" are defined as 1, "T2" are defined as 2, and so on.

We distinguish Derived.AJCC.N..7th.ed..2010.2015.variable. According to its degree, where the degree "N1", "N1a", "N1b", "N1c", "N1NOS" are defined as 1, the degree "N2", "N2a", "N2b", "N2NOS" are defined as 2, and so on.

We define race and marital status separately. The race is divided into four 01 variables, black, white, Asian and Indian, which respectively represent whether they are black, white, Asian and Indian. We separate the marriage state into divorced, married, Separated, single, unmarried, widowed and six 01 variables, respectively indicating whether the 6 different marriage states are satisfied. Seer.historic.stage.a.. 1973.2015. Variable is divided into four 01 variables, including "distant", "stationary", "regional" and "Unstaged", which respectively indicated whether these four conditions were satisfied or not. RX. Summ... Surg.Prim.Site.. variable is divided into five 01 variables according to their size: "RX0", "RX10_19", "RX20_80", "RX90" and "Rx99" to distinguish the position of their size.

It should be noted that since the survival month in the data refers to the time from diagnosis to death, we defined the new survival month as the average of the survival month in the data plus the actual onset date of each tumor stage from diagnosis. The goal is to avoid bias in predicting missing values.

Finally, we re-coded the code.to.site. recode variable, defined it as a classification variable from 1 to 24 in alphabetical order, and defined "Alive" as 0.

Finally, the coded data we obtained were survival month, survival state and 42 factor variables. Due to too many variables, principal component analysis was adopted for 42 variables to extract the main changes and reduce dimension.

## 2.2 Principal component analysis

Principal component analysis, also known as principal component analysis, K-L transform, is a common method in feature extraction methods. When using statistical analysis methods to study multivariate models, too many variables will increase the complexity of the model. It's natural to want more information with fewer variables. In many cases, there is a certain correlation between variables. When there is a certain correlation between two variables, it can be interpreted that the two variables reflect the information of the model has a certain overlap. Principal component analysis is to delete redundant variables (closely related variables) and establish as few new variables as possible, so that these new variables are not correlated in pairs, and these new variables reflect the information of the subject as much as possible to maintain the original information.

The statistical method tries to recombine the original variables into a group of new comprehensive variables unrelated to each other, and at the same time can take out several less comprehensive variables according to the actual needs to reflect the information of the original variables as much as possible. This method is called principal component, which is also a method used for dimensionality reduction in mathematics.

When R was used for principal component analysis, we first made a gravel map, as shown in Figure 2. According to the results in Figure 1, we chose to generate 10 principal component factors instead of 42 variables.

Figure 1. Main component lithotripsy

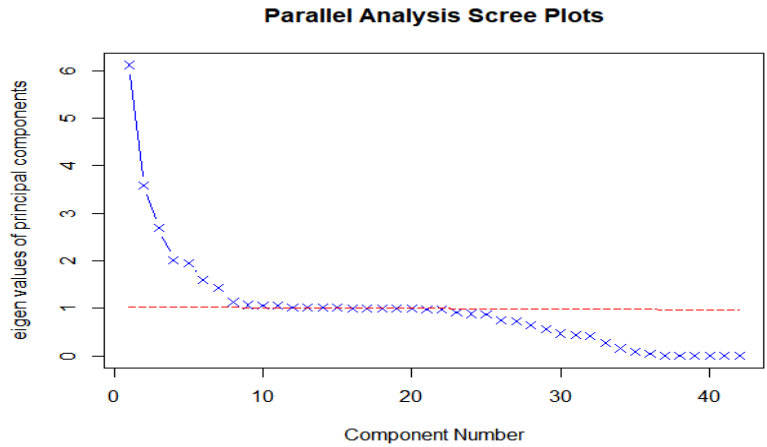

Table 1 Results of principal components

|  | RC1 | RC2 | RC7 | RC3 | RC4 | RC5 | RC6 | RC8 | RC10 | RC9 |
|---|---|---|---|---|---|---|---|---|---|---|
| SS loadings | 4.56 | 3.81 | 3.35 | 2.23 | 1.82 | 1.58 | 1.53 | 1.21 | 1.12 | 1.09 |
| Proportion Var | 0.11 | 0.09 | 0.08 | 0.05 | 0.04 | 0.04 | 0.04 | 0.03 | 0.03 | 0.03 |
| Cumulative Var | 0.11 | 0.20 | 0.28 | 0.33 | 0.38 | 0.41 | 0.45 | 0.51 | 0.54 | 0.57 |

As can be seen from the figure, the first line of Table 1 shows the eigenvalues of each principal component, which all exceed 1. The second line of Table 1 shows the degree of variance explanation of each principal component factor for 42 variables, and the third line shows the degree of cumulative variance explanation. It can be seen from the third line of Table 1 that the 10 principal component factors explain 57% of the variance changes cumulatively and can be added to the model on behalf of 42 variables.

## 2.3 Endpoint event correspondence

According to the survival status and cause of death of the patients, the study divided the end point events into multiple types :(1) The disease of concerned events, that is, the patient died during the follow-up period, and the cause of death was the disease of concern; (2) Competitive events, i.e. death occurred during the follow-up period, but the cause of death was other than the disease concerned, such as death from other cancers while the patient was concerned with heart disease; (3) Deletion, that is, the case was always alive during the follow-up period without death. In Cox proportional risk regression model analysis, the competitive events are not considered, and the competitive events are not considered in the competitive risk model. Traditional survival analysis, such as COX regression, is generally concerned with only one end event (the outcome of interest to the investigator). The individuals who died before recurrence, those who were lost to follow-up and those who did not have recurrence were all treated as Censored Data, requiring that the individual deletion situation was independent of the individual endpoint event, and there was no competition risk for the outcome. In other words, only the death or survival of individuals due to endpoint event A is considered, which is not suitable for data analysis of multiple diseases.

Competing risk model refers to the existence of a certain known event in the observation queue that may affect the occurrence probability of another event or completely hinder its occurrence, so the former and the latter are considered to have competitive risks. As with many diseases, individuals who die from one disease will not die again from other diseases, so it is considered that there is a competitive risk.

In this model we considered 25 different diseases, and give each disease a different index. Five diseases with the largest number of death cases were selected as the concerned diseases of the competitive risk model, and the other diseases were analyzed as competitive risk events in turn. Specific analysis of diseases is shown in the following table.

Table 2 diseases studied

| Disease | The case number | Disease index |
| --- | --- | --- |
| Colon excluding Rectum | 27515 | 7 |
| Rectum and Rectosigmoid Junction | 6450 | 19 |
| Diseases of Heart | 4698 | 24 |
| Other Cause of Death | 2819 | 15 |
| Chronic Obstructive Pulmonary Disease and Allied Cond | 1062 | 6 |

## 2.4 R language implementation

The "Cuminc" function in the "CMPRSK" package of R 4.0.5 software was used to draw the cumulative risk curve, the "CRR" function was used to build the fine-gray competitive risk model, and the "predict" function was used to predict the risk probability.

# 3. Competitive risk model

## 3.1 Fine-gray competitive risk model

Competing Risk Model: refers to the existence of a certain known event in the observation queue that may affect the occurrence probability of another event or completely hinder its occurrence, so the former and the latter are considered to have competitive risks. As with many diseases, individuals who die from one disease will not die again from other diseases, so it is considered that there is a competitive risk.

Competitive risk events: There may be multiple outcome events in the study, some of which will prevent the occurrence of the events of interest or affect the probability of their occurrence. Each outcome event forms a "competitive" relationship and competes with each other. For example, death and dialysis in patients with chronic kidney disease, death and other causes of death in patients with myocardial infarction, death and secondary malignancy in patients with germ cell carcinoma, and postoperative death and pulmonary vein obstruction in patients with congenital heart disease are at competitive risk.

The expression of fine-Gray competitive risk model is:

$$\lambda_k^*(t|X) = \lambda_{k.0}^*(t)\exp(\beta_k^*)^T X$$

The left-hand side of the equation shows the risk calculated using the competitive risk model. λ(t) represents the baseline risk, X is a covariable, and β is a coefficient.
λ(t) can also be expressed as:

$$\lambda_k^*(t|X) = \lim_{\Delta t \to 0} \frac{P(t \leq T < t+\Delta t, D=k | T \geq t U(T \leq t \cap D \neq 1), X)}{\Delta t}$$

It means the probability of death of an individual without ending K occuring before time t, and $\Delta t$ represents unit time.

That is:

$$\lambda_k^*(t|X) = f_k(t)/(1-F_k(t))$$

Where K is the individual and X is the characteristic of the individual, which is our independent variable. On the left hand side is the probability of the sudden end point event that we want to calculate for the individual at time t. $f_k(t)$ represents the probability density function, that is, the time point probability estimate of the occurrence of a terminal event at time t. $F_k(t)$ is a probability distribution function, representing the cumulative probability of occurrence of terminal events up to time T. Then 1-$F_k(t)$ is the probability that no terminal event has occurred until time T. The meaning of the formula is the probability of death of an individual who has no end event K occurring before time t, and the end event K occurring at time t.

In the fine-Gray competitive risk model, the individuals with competitive outcome before time t still exist in the risk set at time t. Let v be the time when the outcome k occurs. For individuals with competitive risk events, v=∞, that is, the outcome K can never occur. Therefore, the fine-Gray competitive risk model well considers the influence of competitive events on the probability of outcome.

The figure below shows a cumulative risk curve for all diseases for one influence factor. In the legend, the first column 0 indicates that the individual is still alive, while 1 indicates that the individual has died. The indices 1, 2, 3.. in the second column represent different diseases. When there are many related diseases, COX survival analysis can only treat single death case and neglect all the other possible diseases, which is obviously not in line with reality. Therefore, it is more realistic to use a competitive risk model at this time.

Figure 2 cumulative risk curve

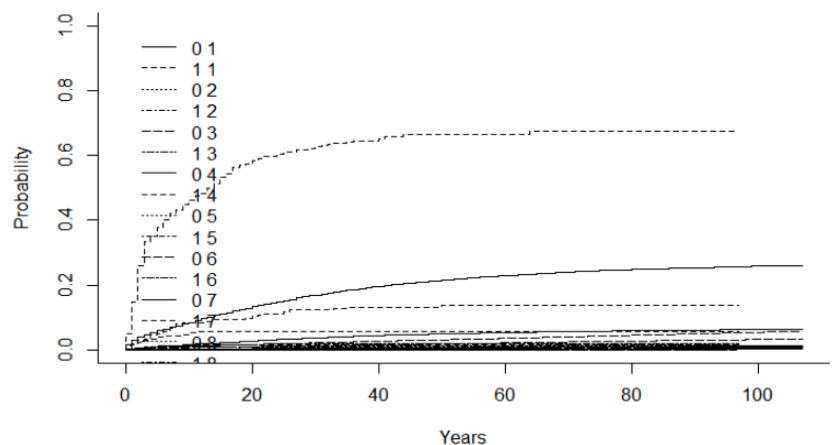

## 3.2 Results

### 3.2.1 Results of Fine Gray competitive risk model for Colon exclusive Rectum disease

When the competitive events were controlled, the Fine Gray competitive risk model results showed that all principal component factors except RC6 were factors affecting the survival of Colon exclusive Rectum patients, with statistically significant differences (P<0.05). In Colon exclusive Rectum patients, the risk of death was 1.099 times higher for each 1 increase in RC1 (95%CI: 1.096-1.103, P<0.001) and 0.992 times higher for each 1 increase in RC2 (95%CI: 1.096-1.103, P<0.001). 0.988-0.996, P<0.001), the risk of death was 0.832 times higher for each 1 increase in RC3 (95%CI: 0.825-0.839, P<0.001), and 1.119 times higher for each 1 increase in RC4 (95%CI: 1.112-1.127, P<0.001), each 1 increase in RC5 was associated with a 0.984 fold increased risk of death (95%CI: 0.977-0.992, P<0.001), and each 1 increase in RC7 was associated with a 1.256 fold increased risk of death (95%CI: 1.248-1.263, P<0.001), the risk of death was 0.982 times higher for each 1 increase in RC8 (95%CI: 0.971-0.992, P<0.001), and 1.025 times higher for each 1 increase in RC9 (95%CI: 1.015 to 1.036, P<0.001), the risk of death was 1.11-fold for each 1 increase in RC10 (95%CI: 1.097 to 1.123, P<0.001).

However, there was no significant difference in the effect of RC6 on the survival of Colon excluding Rectum patients (HR=0.994, 95%CI: 0.985-1.002, P=0.140). The results of Fine Gray competitive risk model are shown in Table 3.

Table 3 Results of Fine Gray competitive risk model of Colon exclusive Rectum disease

|  | exp(coef) | 2.50% | 97.50% | z | p-value |
| --- | --- | --- | --- | --- | --- |
| RC1 | 1.099 | 1.096 | 1.103 | 54.64 | 0.00E+0.0 |
| RC2 | 0.992 | 0.988 | 0.996 | -3.94 | 8.30E-0.5 |
| RC3 | 0.832 | 0.825 | 0.839 | -43.56 | 0.00E+0.0 |
| RC4 | 1.119 | 1.112 | 1.127 | 31.81 | 0.00E+0.0 |
| RC5 | 0.984 | 0.977 | 0.992 | -4.02 | 5.90E-0.5 |
| RC6 | 0.994 | 0.985 | 1.002 | -1.48 | 1.40E-0.1 |
| RC7 | 1.256 | 1.248 | 1.263 | 73.28 | 0.00E+0.0 |
| RC8 | 0.982 | 0.971 | 0.992 | -3.41 | 6.40E-0.4 |
| RC9 | 1.025 | 1.015 | 1.036 | 5 | 5.80E-0.7 |
| RC10 | 1.11 | 1.097 | 1.123 | 17.71 | 0.00E+0.0 |

## 3.2.2 Results of Fine Gray competitive risk model for Rectum and Rectosigmoid Junction disease

When the competitive events were controlled, the results of Fine Gray competitive risk model showed that all principal component factors except RC8 were factors influencing the survival of patients with Rectum and Rectosigmoid Junction, and the differences were statistically significant (P<0.05). In patients with Rectum and Rectosigmoid Junction, the risk of death was 1.021 times higher for each 1 increase in RC1 (95% CI: 1.015-1.028, P<0.001), each 1 increase in RC2 was associated with a 1.122-fold increased risk of death (95%CI: 1.115-1.129, P<0.001), and each 1 increase in RC3 was associated with a 1.292-fold increased risk of death (95%CI: 1.28 to 1.305, P<0.001), the risk of death was 1.057 times higher for each 1 increase in RC4 (95%CI: 1.042-1.073, P<0.001), and 1.016 times higher for each 1 increase in RC5 (95%CI:

1.016). 1.001 to 1.031, P<0.001), each 1 increase in RC6 was associated with a 1.029-fold increased risk of death (95%CI: 1.013-1.045, P<0.001), and each 1 increase in RC7 was associated with a 1.097 fold increased risk of death (95%CI: 1.085-1.109, P<0.001), each 1 increase in RC9 was associated with a 0.946-fold increased risk of death (95%CI: 0.937-0.956, P<0.001), and each 1 increase in RC10 was associated with a 1.015-fold increased risk of death (95%CI: 0.999~1.032, P<0.001).

However, RC8 had no significant difference in the survival status of Rectum and Rectosigmoid Junction patients (HR=1.005, 95%CI: 0.986-1.025, P=0.590). The results of the Fine Gray competitive risk model are shown in Table 4.

Table 4 Results of Fine Gray competitive risk model for Rectum and Rectosigmoid Junction disease

|      | exp(coef) | 2.50% | 97.50% | z       | p-value   |
|------|-----------|-------|--------|---------|-----------|
| RC1  | 1.021     | 1.015 | 1.028  | 6.334   | 2.40E-1.0 |
| RC2  | 1.122     | 1.115 | 1.129  | 35.462  | 0.00E+0.0 |
| RC3  | 1.292     | 1.28  | 1.305  | 52.526  | 0.00E+0.0 |
| RC4  | 1.057     | 1.042 | 1.073  | 7.325   | 2.40E-1.3 |
| RC5  | 1.016     | 1.001 | 1.031  | 2.031   | 4.20E-0.2 |
| RC6  | 1.029     | 1.013 | 1.045  | 3.594   | 3.30E-0.4 |
| RC7  | 1.097     | 1.085 | 1.109  | 16.572  | 0.00E+0.0 |
| RC8  | 1.005     | 0.986 | 1.025  | 0.532   | 5.90E-0.1 |
| RC9  | 0.946     | 0.937 | 0.956  | -10.604 | 0.00E+0.0 |
| RC10 | 1.015     | 0.999 | 1.032  | 1.784   | 7.40E-0.2 |

## 3.2.3 Results of The Fine Gray competitive risk model for Diseases of Heart

When competitive events were controlled, the results of Fine Gray competitive risk model showed that all principal component factors except RC7 were influential factors for the survival of Diseases of Heart patients, with statistically significant differences (ALL P<0.05). In Diseases of Heart patients, the risk of death was 0.915 times higher for each 1 increase in RC1 (95%CI: 0.903-0.928, P<0.001), and 1.013 times higher for each 1 increase in RC2 (95%CI: 1.002 to 1.024, P<0.001), each 1 increase in RC3 was associated with a 0.914-fold increased risk of death (95%CI: 0.895 to 0.933, P<0.001), and each 1 increase in RC4 was associated with a 1.258 fold increased risk of death (95%CI: 1.242-1.275, P<0.001), the risk of death was 0.92 times higher for each 1 increase in RC5 (95%CI: 0.901-0.939, P<0.001), and 0.938 times higher for each 1 increase in RC6 (95%CI: 0.914-0.963, P<0.001), each 1 increase in RC8 was associated with a 0.931-fold increased risk of death (95%CI: 0.902-0.962, P<0.001), and each 1 increase in RC9 was associated with a 1.169 fold increased risk of death (95%CI: 1.148 to 1.19, P<0.001), each 1 increase in RC10 was associated with a 1.115-fold increased risk of death (95%CI: 1.092 to 1.139, P<0.001).

However, there was no statistical significance in the influence of RC7 on the survival of Diseases of Heart patients (HR=0.991, 95%CI: 0.902~1.002, P=0.130). The results of the Fine Gray competitive risk model are shown in Table 5.

Table 5 Results of Fine Gray competitive risk model for Diseases of Heart Diseases

|  | exp(coef) | 2.50% | 97.50% | z | p-value |
| --- | --- | --- | --- | --- | --- |
| RC1 | 0.915 | 0.903 | 0.928 | -12.82 | 0.00E+0.0 |
| RC2 | 1.013 | 1.002 | 1.024 | 2.24 | 2.50E-0.2 |
| RC3 | 0.914 | 0.895 | 0.933 | -8.51 | 0.00E+0.0 |
| RC4 | 1.258 | 1.242 | 1.275 | 33.95 | 0.00E+0.0 |
| RC5 | 0.92 | 0.901 | 0.939 | -7.85 | 4.00E-1.5 |
| RC6 | 0.938 | 0.914 | 0.963 | -4.78 | 1.70E-0.6 |
| RC7 | 0.991 | 0.98 | 1.002 | -1.53 | 1.30E-0.1 |
| RC8 | 0.931 | 0.902 | 0.962 | -4.34 | 1.40E-0.5 |
| RC9 | 1.169 | 1.148 | 1.19 | 16.96 | 0.00E+0.0 |
| RC10 | 1.115 | 1.092 | 1.139 | 10.09 | 0.00E+0.0 |

### 3.2.4 Results of Fine Gray competitive Risk model for other diseases

When the competitive events were controlled, the results of Fine Gray competitive risk model showed that the principal component factors except RC2, RC7 and RC8 were all influential factors for the survival of patients with Other Cause of Death, and the differences were statistically significant (P<0.05). In Other Cause of Death patients, the risk of Death was 0.939 times higher for each 1 increase in RC1 (95%CI: 0.925-0.954, P<0.001), and 0.914 times higher for each 1 increase in RC3 (95%CI: 0.89 to 0.938, P<0.001), the risk of death was 1.229 times higher for each 1 increase in RC4 (95%CI: 1.208 to 1.25, P<0.001), and 0.908 times higher for each 1 increase in RC5 (95%CI: 1.208 to 1.25, P<0.001). 0.883 to 0.933, P<0.001), the risk of death was 0.961 times higher for each 1 increase in RC6 (95%CI: 0.93-0.992, P<0.001), and 1.126 times higher for each 1 increase in RC9 (95%CI: 0.93-0.992, P<0.001). 1.101 to 1.152, P<0.001), the risk of death was 1.068 times higher for each 1 increase in RC10 (95%CI: 1.038 to 1.099, P<0.001).

However, there was no significant difference in the impact of RC2 on the survival of patients with Other Cause of Death (HR=1.004, 95%CI: 0.99-1.018, P=0.580). There was no significant difference in the impact of RC7 on the survival of patients with Other Cause of Death (HR=1.01, 95%CI: 0.995-1.024, P=0.190). RC8 had no significant difference in the survival status of patients with Other Cause of Death (HR=0.983, 95%CI: 0.948-1.019, P=0.340).The results of Fine Gray competitive risk model are shown in Table 6.

Table 6 Results of Fine Gray competitive risk model of Other Cause of Death Diseases

|      | exp(coef) | 2.50% | 97.50% | z      | p-value    |
|------|-----------|-------|--------|--------|------------|
| RC1  | 0.939     | 0.925 | 0.954  | -8.114 | 4.40E-1.6  |
| RC2  | 1.004     | 0.99  | 1.018  | 0.553  | 5.80E-0.1  |
| RC3  | 0.914     | 0.89  | 0.938  | -6.652 | 2.90E-1.1  |
| RC4  | 1.229     | 1.208 | 1.25   | 23.818 | 0.00E+0.0  |
| RC5  | 0.908     | 0.883 | 0.933  | -6.854 | 7.20E-1.2  |
| RC6  | 0.961     | 0.93  | 0.992  | -2.436 | 1.50E-0..2 |
| RC7  | 1.01      | 0.995 | 1.024  | 1.301  | 1.90E-0.1  |
| RC8  | 0.983     | 0.948 | 1.019  | -0.945 | 3.40E-0.1  |
| RC9  | 1.126     | 1.101 | 1.152  | 10.211 | 0.00E+0.0  |
| RC10 | 1.068     | 1.038 | 1.099  | 4.469  | 7.90E-0.6  |

### 3.2.5 Results of the Fine Gray competitive risk model for Chronic Obstructive Pulmonary Disease and Allied Cond Disease

The results of the Fine Gray competitive risk model showed that all the principal component factors except RC6 and RC7 were factors influencing the survival of patients with Chronic Obstructive Pulmonary Disease and Allied Cond when the competitive events were controlled. The differences were statistically significant (P<0.05). In patients with Chronic Obstructive Pulmonary Disease and Allied Cond, the risk of death increased by 0.889 times for each 1 increase in RC1 (95% CI: 0.863-0.916, P<0.001), the risk of death was 1.023 times higher for each 1 increase in RC2 (95%CI: 1.002-1.045, P<0.001), and 0.918 times higher for each 1 increase in RC3 (95%CI: 0.881-0.957, P<0.001), the risk of death was 1.242 times higher for each 1 increase in RC4 (95%CI: 1.21-1.276, P<0.001), and 0.849 times higher for each 1 increase in RC5 (95%CI: 1.21-1.276, P<0.001). 0.805-0.895, P<0.001), each 1 increase in RC8 was associated with a 1.072-fold increased risk of death (95%CI: 1.017-1.129, P<0.001), and each 1 increase in RC9 was associated with a 1.205 fold increased risk of death (95%CI: 1.165 to 1.245, P<0.001), the risk of death was 1.088 times higher for each 1 increase in RC10 (95%CI: 1.043 to 1.134, P<0.001).

However, RC6 had no significant effect on the survival of patients with Chronic Obstructive Pulmonary Disease and Allied Cond (HR=0.97, 95%CI: 0.918-1.024, P=0.270). RC7 had no significant effect on the survival of patients with Chronic Obstructive Pulmonary Disease and Allied Cond (HR=0.998, 95%CI: 0.976~1.021, P=0.880). The results of the Fine Gray competitive risk model are shown in Table 7.

Table 7 Results of the Chronic Obstructive Pulmonary Disease and Allied Cond Disease competitive risk model

|      | exp(coef) | 2.50% | 97.50% | z      | p-value   |
|------|-----------|-------|--------|--------|-----------|
| RC1  | 0.889     | 0.863 | 0.916  | -7.771 | 7.80E-1.5 |
| RC2  | 1.023     | 1.002 | 1.045  | 2.161  | 3.10E-0.2 |
| RC3  | 0.918     | 0.881 | 0.957  | -4.017 | 5.90E-0.5 |
| RC4  | 1.242     | 1.21  | 1.276  | 16.041 | 0.00E+0.0 |
| RC5  | 0.849     | 0.805 | 0.895  | -6.09  | 1.10E-0.9 |
| RC6  | 0.97      | 0.918 | 1.024  | -1.096 | 2.70E-0.1 |
| RC7  | 0.998     | 0.976 | 1.021  | -0.155 | 8.80E-0.1 |
| RC8  | 1.072     | 1.017 | 1.129  | 2.59   | 9.60E-0.3 |
| RC9  | 1.205     | 1.165 | 1.245  | 11.056 | 0.00E+0.0 |
| RC10 | 1.088     | 1.043 | 1.134  | 3.915  | 9.00E-0.5 |

## 3.3 Summarize

Combining the results of several diseases, RC4 and RC10 have adverse effects on all five diseases, and increasing RC4 and RC10 will lead to an increased risk of death. RC1,RC3 and RC5 have a good effect on diseases outside Rectum and Rectosigmoid Junction, and individuals with higher RC1,RC3 and RC5 have a relatively low risk of death. RC2 had an adverse effect on diseases other than Colon exclusive Rectum, resulting in an increased risk of individual death. RC9 has adverse effects on diseases other than Rectum and Rectosigmoid Junction, resulting in an increased risk of individual death. However, RC6,RC7 and RC8 do not have significant effects on multiple diseases, so it can be considered that they have no impact on individual mortality risk.

## 3.4 When not using principal component analysis

In addition, we also try to analyze the data by competitive risk model without applying principal component analysis. Theoretically, the use of original data will enable us to more intuitively observe the impact of each variable on the risk of death from a certain disease, while the results of principal component analysis need to consider which variables are composed of each principal component factor, and then further judge the impact of each variable. So in theory do not use the principal component to make risk factors determine which factor is more convenient, using principal component will be able to see the comprehensive effect of various factors, and on the basis of the results of principal components can also see what factors are the changes between comparing similar (if the principal component factor 1 mainly consists of five variables, shows that there are similarities between the five variables change, High correlation). In addition, the use of original data will lead to large amount of data and excessive classification, which will easily lead to errors in the calculation process. Therefore, after reducing some variables, the data without principal component is used for competitive risk model analysis.

### 3.4.1 The competitive risk model calculation for Diseases of Heart is shown in the figure below

```
                                              coef exp(coef) se(coef)        z p-value
Age.recode.with.single.ages.and.100.       0.070791     1.073 0.001445  49.0005 0.0e+00
Sex                                        0.519885     1.682 0.031570  16.4676 0.0e+00
Grade..thru.2017.                         -0.001785     0.998 0.024504  -0.0729 9.4e-01
Chemotherapy.recode..yes..no.unk.         -0.367285     0.693 0.042975  -8.5465 0.0e+00
SEER.Combined.Mets.at.DX.bone..2010..     -1.310432     0.270 0.502562  -2.6075 9.1e-03
SEER.Combined.Mets.at.DX.brain..2010..     0.348448     1.417 0.465854   0.7480 4.5e-01
SEER.Combined.Mets.at.DX.liver..2010..    -0.174523     0.840 0.119487  -1.4606 1.4e-01
SEER.Combined.Mets.at.DX.lung..2010..     -0.314272     0.730 0.146512  -2.1450 3.2e-02
Derived.AJCC.Stage.Group..7th.ed..2010.2015. -0.042093  0.959 0.025530  -1.6488 9.9e-02
Derived.AJCC.T..7th.ed..2010.2015.        -0.020227     0.980 0.015865  -1.2750 2.0e-01
Derived.AJCC.N..7th.ed..2010.2015.        -0.068473     0.934 0.026871  -2.5482 1.1e-02
Derived.AJCC.M..7th.ed..2010.2015.        -0.486804     0.615 0.112653  -4.3213 1.6e-05
Regional.nodes.positive..1988..            0.000285     1.000 0.000447   0.6380 5.2e-01
white                                     -0.176598     0.838 0.046970  -3.7598 1.7e-04
indian                                    -0.180894     0.835 0.202282  -0.8943 3.7e-01
asian                                     -0.595055     0.552 0.076214  -7.8076 5.8e-15
married                                   -0.282851     0.754 0.037956  -7.4521 9.2e-14
separated                                 -0.050227     0.951 0.158694  -0.3165 7.5e-01
unmarried                                 -0.980127     0.375 0.566920  -1.7289 8.4e-02
widowed                                   -0.077873     0.925 0.044089  -1.7663 7.7e-02
```

As can be seen from the figure, the single factor test (Z test) of most variables is significant at the significance level of 5% (i.e., the corresponding p-value<0.05). That is to say, most variables have a significant impact on the dependent variable (that is, the regression coefficient for the dependent variable is significantly not 0), and a few variables have no significant impact (that is, the regression coefficient for the dependent variable may be 0, that is, there may be no statistical impact). The regression coefficient is divided into COEF and EXP (COEF), and exp(COEF) is what we need here, which represents the risk impact of independent variable on dependent variable. When exp(COEF)>1, it means that the independent variable is a risk factor of the dependent variable. For example, the exponential regression coefficient of Age in the result is 1.073, that is to say, the older the Age, the greater the risk of death. For each 1-year increase in age, the risk of death increased 1.073 times. Similarly, we can intuitively observe the impact of various variables on the risk of death.

Look at the figure below, which contains the value of the exponential regression coefficient and the 95% confidence interval (from 2.5% to 97.5%), illustrating the range in which the 95% probability of the regression coefficient is likely to be.

```
                                               exp(coef) exp(-coef)  2.5%  97.5%
Age.recode.with.single.ages.and.100.              1.073      0.932  1.070  1.076
Sex                                               1.682      0.595  1.581  1.789
Grade..thru.2017.                                 0.998      1.002  0.951  1.047
Chemotherapy.recode..yes..no.unk.                 0.693      1.444  0.637  0.753
SEER.Combined.Mets.at.DX.bone..2010..             0.270      3.708  0.101  0.722
SEER.Combined.Mets.at.DX.brain..2010..            1.417      0.706  0.569  3.531
SEER.Combined.Mets.at.DX.liver..2010..            0.840      1.191  0.665  1.061
SEER.Combined.Mets.at.DX.lung..2010..             0.730      1.369  0.548  0.973
Derived.AJCC.Stage.Group..7th.ed..2010.2015.      0.959      1.043  0.912  1.008
Derived.AJCC.T..7th.ed..2010.2015.                0.980      1.020  0.950  1.011
Derived.AJCC.N..7th.ed..2010.2015.                0.934      1.071  0.886  0.984
Derived.AJCC.M..7th.ed..2010.2015.                0.615      1.627  0.493  0.766
Regional.nodes.positive..1988..                   1.000      1.000  0.999  1.001
white                                             0.838      1.193  0.764  0.919
indian                                            0.835      1.198  0.561  1.241
asian                                             0.552      1.813  0.475  0.640
married                                           0.754      1.327  0.700  0.812
separated                                         0.951      1.052  0.697  1.298
unmarried                                         0.375      2.665  0.124  1.140
widowed                                           0.925      1.081  0.849  1.009
```

## 3.4.2 The calculation results of the competition risk model for Colon exclusive Rectum are shown in the figure below

Look at the figure below, which contains the value of the exponential regression coefficient and the 95% confidence interval (from 2.5% to 97.5%), illustrating the range in which the 95% probability of the regression coefficient is likely to be.

```
                                                  coef exp(coef)  se(coef)      z p-value
Age.recode.with.single.ages.and.100.           0.020121    1.020  0.000561  35.850 0.0e+00
Sex                                            0.008315    1.008  0.013442   0.619 5.4e-01
Grade..thru.2017.                              0.237255    1.268  0.010240  23.170 0.0e+00
Chemotherapy.recode..yes..no.unk.             -0.305914    0.736  0.015991 -19.131 0.0e+00
SEER.Combined.Mets.at.DX.bone..2010..         -0.064557    0.937  0.055260  -1.168 2.4e-01
SEER.Combined.Mets.at.DX.brain..2010..         0.197611    1.218  0.102254   1.933 5.3e-02
SEER.Combined.Mets.at.DX.liver..2010..         0.256272    1.292  0.021619  11.854 0.0e+00
SEER.Combined.Mets.at.DX.lung..2010..         -0.013296    0.987  0.025256  -0.526 6.0e-01
Derived.AJCC.Stage.Group..7th.ed..2010.2015.   0.445086    1.561  0.015047  29.580 0.0e+00
Derived.AJCC.T..7th.ed..2010.2015.             0.262789    1.301  0.008562  30.693 0.0e+00
Derived.AJCC.N..7th.ed..2010.2015.             0.238266    1.269  0.009216  25.854 0.0e+00
Derived.AJCC.M..7th.ed..2010.2015.             0.688085    1.990  0.029159  23.597 0.0e+00
Regional.nodes.positive..1988..                0.000958    1.001  0.000197   4.863 1.2e-06
white                                         -0.222645    0.800  0.019062 -11.680 0.0e+00
indian                                        -0.030379    0.970  0.071050  -0.428 6.7e-01
asian                                         -0.307859    0.735  0.028400 -10.840 0.0e+00
married                                       -0.104167    0.901  0.015103  -6.897 5.3e-12
separated                                     -0.039173    0.962  0.058716  -0.667 5.0e-01
unmarried                                     -0.206535    0.813  0.145181  -1.423 1.5e-01
widowed                                       -0.004329    0.996  0.022581  -0.192 8.5e-01

                                               exp(coef) exp(-coef)  2.5%  97.5%
Age.recode.with.single.ages.and.100.              1.020      0.980  1.019  1.021
Sex                                               1.008      0.992  0.982  1.035
Grade..thru.2017.                                 1.268      0.789  1.243  1.293
Chemotherapy.recode..yes..no.unk.                 0.736      1.358  0.714  0.760
SEER.Combined.Mets.at.DX.bone..2010..             0.937      1.067  0.841  1.045
SEER.Combined.Mets.at.DX.brain..2010..            1.218      0.821  0.997  1.489
SEER.Combined.Mets.at.DX.liver..2010..            1.292      0.774  1.238  1.348
SEER.Combined.Mets.at.DX.lung..2010..             0.987      1.013  0.939  1.037
Derived.AJCC.Stage.Group..7th.ed..2010.2015.      1.561      0.641  1.515  1.607
Derived.AJCC.T..7th.ed..2010.2015.                1.301      0.769  1.279  1.323
Derived.AJCC.N..7th.ed..2010.2015.                1.269      0.788  1.246  1.292
Derived.AJCC.M..7th.ed..2010.2015.                1.990      0.503  1.879  2.107
Regional.nodes.positive..1988..                   1.001      0.999  1.001  1.001
white                                             0.800      1.249  0.771  0.831
indian                                            0.970      1.031  0.844  1.115
asian                                             0.735      1.361  0.695  0.777
married                                           0.901      1.110  0.875  0.928
separated                                         0.962      1.040  0.857  1.079
unmarried                                         0.813      1.229  0.612  1.081
widowed                                           0.996      1.004  0.953  1.041
```

The results of z-test and 95% confidence intervals are shown in the figure respectively. The regression coefficient is divided into COEF and EXP (COEF). Exp (COEF) is what we need here, which represents the risk impact of independent variable on dependent variable. When exp(COEF)>1, it means that the independent variable is a risk factor of the dependent variable. According to the figure, age, sex, Grade... Thru. 2017., SEER.Com bined. Mets. Palawan DX. Brian.. 2010.. , SEER.Com bined. Mets. Ats. DX. Liver.. 2010.. , Derived. AJCC. Stage. The Group.. 7th.ed.. 2010.2015., Derived. AJCC. T.. 7th.ed.. 2010.2015., Derived. AJCC. N.. 7th.ed.. 2010.2015., Derived. AJCC. M.. 7th.ed.. 2010.2015., Regional. Nodes. Positive.. 1988.. These factors are risk factors, and their exponential regression coefficients are all greater than 1, indicating that the larger these variables are, the greater the risk of death will be. Chemotherapy.recode.. yes.. No. Unk., SEER.Com bined. Mets. Palawan DX. Ipads.. 2010.. , SEER.Com bined. Mets. Ats. DX. Lung.. 2010.. , White, Indian, Asian, Married, Separated, unmarried, widowed these variables are not risk factors, and their exponential regression coefficients are all less than 1, indicating that the greater these variables are, the lower the risk of death is. If the exponential regression element of age is 1.02, it indicates that the risk of death is 1.02 times higher for every 1 increase in age. Such as SEER.Com bined. Mets. Ats. DX. Ipads.. 2010.. Index of the regression coefficient is 0.937, that SEER.Com bined. Mets. Palawan DX. Ipads.. 2010.. For each 1 increase, the risk of death increased 0.937 times.

## 3.4.3 The calculation results of the competition risk model for Rectum and Rectosigmoid Junction are shown in the figure below

```
                                              coef exp(coef) se(coef)       z p-value
Age.recode.with.single.ages.and.100.        0.0052     1.005 0.001046   4.965 6.9e-07
Sex                                         0.2819     1.326 0.026719  10.550 0.0e+00
Grade..thru.2017.                           0.1016     1.107 0.019843   5.118 3.1e-07
Chemotherapy.recode..yes..no.unk.           0.9012     2.463 0.036161  24.922 0.0e+00
SEER.Combined.Mets.at.DX.bone..2010..       0.2700     1.310 0.074696   3.615 3.0e-04
SEER.Combined.Mets.at.DX.brain..2010..     -0.1971     0.821 0.180850  -1.090 2.8e-01
SEER.Combined.Mets.at.DX.liver..2010..     -0.0310     0.969 0.044479  -0.697 4.9e-01
SEER.Combined.Mets.at.DX.lung..2010..       0.2941     1.342 0.042589   6.905 5.0e-12
Derived.AJCC.Stage.Group..7th.ed..2010.2015. 0.2707    1.311 0.025922  10.443 0.0e+00
Derived.AJCC.T..7th.ed..2010.2015.          0.0592     1.061 0.011563   5.118 3.1e-07
Derived.AJCC.N..7th.ed..2010.2015.          0.0561     1.058 0.015545   3.612 3.0e-04
Derived.AJCC.M..7th.ed..2010.2015.         -0.1667     0.846 0.061328  -2.719 6.6e-03
Regional.nodes.positive..1988..             0.0162     1.016 0.000327  49.570 0.0e+00
white                                       0.2064     1.229 0.041519   4.971 6.7e-07
indian                                      0.3996     1.491 0.131355   3.042 2.3e-03
asian                                       0.3228     1.381 0.055819   5.782 7.4e-09
married                                    -0.2845     0.752 0.028592  -9.951 0.0e+00
separated                                  -0.0375     0.963 0.105439  -0.355 7.2e-01
unmarried                                  -0.1357     0.873 0.219432  -0.618 5.4e-01
widowed                                     0.0485     1.050 0.044753   1.084 2.8e-01

                                             exp(coef) exp(-coef) 2.5% 97.5%
Age.recode.with.single.ages.and.100.             1.005      0.995 1.003 1.007
Sex                                              1.326      0.754 1.258 1.397
Grade..thru.2017.                                1.107      0.903 1.065 1.151
Chemotherapy.recode..yes..no.unk.                2.463      0.406 2.294 2.643
SEER.Combined.Mets.at.DX.bone..2010..            1.310      0.763 1.132 1.517
SEER.Combined.Mets.at.DX.brain..2010..           0.821      1.218 0.576 1.170
SEER.Combined.Mets.at.DX.liver..2010..           0.969      1.031 0.889 1.058
SEER.Combined.Mets.at.DX.lung..2010..            1.342      0.745 1.234 1.459
Derived.AJCC.Stage.Group..7th.ed..2010.2015.     1.311      0.763 1.246 1.379
Derived.AJCC.T..7th.ed..2010.2015.               1.061      0.943 1.037 1.085
Derived.AJCC.N..7th.ed..2010.2015.               1.058      0.945 1.026 1.090
Derived.AJCC.M..7th.ed..2010.2015.               0.846      1.181 0.751 0.955
Regional.nodes.positive..1988..                  1.016      0.984 1.016 1.017
white                                            1.229      0.814 1.133 1.333
indian                                           1.491      0.671 1.153 1.929
asian                                            1.381      0.724 1.238 1.541
married                                          0.752      1.329 0.711 0.796
separated                                        0.963      1.038 0.783 1.184
unmarried                                        0.873      1.145 0.568 1.342
widowed                                          1.050      0.953 0.962 1.146
```

The results of z-test and 95% confidence intervals are shown in the figure respectively.

According to the figure, age, sex, Grade... Thru. 2017., Chemotherapy. The recode.. yes.. No. Unk., SEER.Com bined. Mets. Palawan DX. Ipads.. 2010.. , SEER.Com bined. Mets. Ats. DX. Lung.. 2010.. , Derived. AJCC. Stage. The Group.. 7th.ed.. 2010.2015., Derived. AJCC. T.. 7th.ed.. 2010.2015., Derived. AJCC. N.. 7th.ed.. 2010.2015., Regional. Nodes. Positive.. 1988.. , White, Indian, Asian and Widowed are risk factors, and their exponential regression coefficients are all greater than 1, indicating that the greater these variables are, the greater the risk of death will be. SEER.Combined.Mets.at.DX.brain.. 2010.. SEER.Combined.Mets.at.DX.liver.. 2010.. , Derived. AJCC. M.. 7th.ed.. 2010, 2015., Married, Separated, unmarried, these variables are not risk factors, their exponential regression coefficients are all less than 1, indicating that the larger these variables are, the lower the risk of death. If the exponential regression element of age is 1.005, it means that the risk of death increases 1.005 times for every 1 increase in age. Such as SEER.Com bined. Mets. Ats. DX. Brian.. 2010.. Index of the regression coefficient is 0.821, that SEER.Com bined. Mets. Palawan DX. Brian.. 2010.. For each 1 increase, the risk of death increased by 0.821 times.

## 3.4.4 The calculation results of the Other Cause of Death competitive risk model are shown in the figure below

```
                                               coef exp(coef) se(coef)       z p-value
Age.recode.with.single.ages.and.100.        0.051368     1.053  0.00181 28.4296 0.0e+00
Sex                                         0.143560     1.154  0.04050  3.5444 3.9e-04
Grade..thru.2017.                           0.000626     1.001  0.03221  0.0194 9.8e-01
Chemotherapy.recode..yes..no.unk.          -0.486058     0.615  0.05333 -9.1143 0.0e+00
SEER.Combined.Mets.at.DX.bone..2010..      -0.072153     0.930  0.32956 -0.2189 8.3e-01
SEER.Combined.Mets.at.DX.brain..2010..     -0.354964     0.701  0.71833 -0.4942 6.2e-01
SEER.Combined.Mets.at.DX.liver..2010..      0.131602     1.141  0.14050  0.9367 3.5e-01
SEER.Combined.Mets.at.DX.lung..2010..      -0.313761     0.731  0.16812 -1.8663 6.2e-02
Derived.AJCC.Stage.Group..7th.ed..2010.2015. 0.020189    1.020  0.03250  0.6213 5.3e-01
Derived.AJCC.T..7th.ed..2010.2015.          0.022486     1.023  0.02044  1.1000 2.7e-01
Derived.AJCC.N..7th.ed..2010.2015.         -0.014819     0.985  0.03216 -0.4608 6.4e-01
Derived.AJCC.M..7th.ed..2010.2015.         -0.591178     0.554  0.13679 -4.3218 1.5e-05
Regional.nodes.positive..1988..            -0.000693     0.999  0.00060 -1.1539 2.5e-01
white                                       0.029267     1.030  0.06320  0.4631 6.4e-01
indian                                      0.168862     1.184  0.23313  0.7243 4.7e-01
asian                                      -0.482662     0.617  0.10169 -4.7463 2.1e-06
married                                    -0.267892     0.765  0.04768 -5.6181 1.9e-08
separated                                  -0.091977     0.912  0.20333 -0.4524 6.5e-01
unmarried                                   0.006019     1.006  0.44875  0.0134 9.9e-01
widowed                                    -0.150049     0.861  0.05696 -2.6344 8.4e-03

                                             exp(coef) exp(-coef)  2.5% 97.5%
Age.recode.with.single.ages.and.100.             1.053      0.950 1.049 1.056
Sex                                              1.154      0.866 1.066 1.250
Grade..thru.2017.                                1.001      0.999 0.939 1.066
Chemotherapy.recode..yes..no.unk.                0.615      1.626 0.554 0.683
SEER.Combined.Mets.at.DX.bone..2010..            0.930      1.075 0.488 1.775
SEER.Combined.Mets.at.DX.brain..2010..           0.701      1.426 0.172 2.866
SEER.Combined.Mets.at.DX.liver..2010..           1.141      0.877 0.866 1.502
SEER.Combined.Mets.at.DX.lung..2010..            0.731      1.369 0.526 1.016
Derived.AJCC.Stage.Group..7th.ed..2010.2015.     1.020      0.980 0.957 1.087
Derived.AJCC.T..7th.ed..2010.2015.               1.023      0.978 0.983 1.065
Derived.AJCC.N..7th.ed..2010.2015.               0.985      1.015 0.925 1.049
Derived.AJCC.M..7th.ed..2010.2015.               0.554      1.806 0.423 0.724
Regional.nodes.positive..1988..                  0.999      1.001 0.998 1.000
white                                            1.030      0.971 0.910 1.165
indian                                           1.184      0.845 0.750 1.870
asian                                            0.617      1.620 0.506 0.753
married                                          0.765      1.307 0.697 0.840
separated                                        0.912      1.096 0.612 1.359
unmarried                                        1.006      0.994 0.417 2.424
widowed                                          0.861      1.162 0.770 0.962
```

The results of z-test and 95% confidence intervals are shown in the figure respectively. According to the figure, age, sex, Grade... Thru. 2017., SEER.Com bined. Mets. Palawan DX.

Liver.. 2010.. , Derived. AJCC. Stage. The Group.. 7th.ed.. 2010.2015., Derived. AJCC. T.. 7th.ed.. 2010.2015., White, knight these factors are risk factors, their exponential regression coefficients are all greater than 1, indicating that the larger these variables are, the greater the risk of death. Other variables are not risk factors, and their exponential regression coefficients are all less than 1, indicating that the larger these variables are, the lower the risk of death is. If the exponential regression element of age is 1.053, it indicates that the risk of death is 1.053 times higher for every 1 increase in age. Such as SEER.Com bined. Mets. Ats. DX. Ipads.. 2010.. Index of the regression coefficient is 0.93, that SEER.Com bined. Mets. Palawan DX. Ipads.. 2010.. For every 1 increase, the risk of death increased by 0.93 times.

## 3.4.5 The results of our competitive risk model for Chronic Obstructive Pulmonary Disease and Allied Cond are shown in the figure below

```
                                              coef  exp(coef)  se(coef)       z p-value
Age.recode.with.single.ages.and.100.       0.05750   1.059182  0.002521  22.8091 0.0e+00
Sex                                        0.22848   1.256684  0.066067   3.4583 5.4e-04
Grade..thru.2017.                         -0.01746   0.982696  0.053751  -0.3247 7.5e-01
Chemotherapy.recode..yes..no.unk.         -0.48933   0.613036  0.091218  -5.3644 8.1e-08
SEER.Combined.Mets.at.DX.bone..2010..      0.62534   1.868885  0.481029   1.3000 1.9e-01
SEER.Combined.Mets.at.DX.brain..2010..    -8.19274   0.000277  0.192593 -42.5392 0.0e+00
SEER.Combined.Mets.at.DX.liver..2010..    -0.44592   0.640236  0.263130  -1.6947 9.0e-02
SEER.Combined.Mets.at.DX.lung..2010..      0.12461   1.132710  0.301415   0.4134 6.8e-01
Derived.AJCC.Stage.Group..7th.ed..2010.2015. 0.00821 1.008241  0.053429   0.1536 8.8e-01
Derived.AJCC.T..7th.ed..2010.2015.         0.00302   1.003030  0.033802   0.0895 9.3e-01
Derived.AJCC.N..7th.ed..2010.2015.        -0.13441   0.874233  0.056609  -2.3743 1.8e-02
Derived.AJCC.M..7th.ed..2010.2015.        -0.75244   0.471214  0.235958  -3.1889 1.4e-03
Regional.nodes.positive..1988..            0.00114   1.001136  0.000878   1.2932 2.0e-01
white                                      0.35788   1.430299  0.115989   3.0855 2.0e-03
indian                                    -0.18194   0.833652  0.513240  -0.3545 7.2e-01
asian                                     -0.39777   0.671820  0.186256  -2.1356 3.3e-02
married                                   -0.53014   0.588520  0.076249  -6.9528 3.6e-12
separated                                 -0.20598   0.813847  0.338438  -0.6086 5.4e-01
unmarried                                 -0.78704   0.455192  1.005140  -0.7830 4.3e-01
widowed                                   -0.28354   0.753113  0.091577  -3.0962 2.0e-03

                                          exp(coef) exp(-coef)    2.5%    97.5%
Age.recode.with.single.ages.and.100.       1.059182      0.944 1.05396 1.064428
Sex                                        1.256684      0.796 1.10405 1.430415
Grade..thru.2017.                          0.982696      1.018 0.88444 1.091873
Chemotherapy.recode..yes..no.unk.          0.613036      1.631 0.51267 0.733046
SEER.Combined.Mets.at.DX.bone..2010..      1.868885      0.535 0.72800 4.797727
SEER.Combined.Mets.at.DX.brain..2010..     0.000277   3614.621 0.00019 0.000404
SEER.Combined.Mets.at.DX.liver..2010..     0.640236      1.562 0.38226 1.072300
SEER.Combined.Mets.at.DX.lung..2010..      1.132710      0.883 0.62741 2.044957
Derived.AJCC.Stage.Group..7th.ed..2010.2015. 1.008241    0.992 0.90800 1.119550
Derived.AJCC.T..7th.ed..2010.2015.         1.003030      0.997 0.93873 1.071732
Derived.AJCC.N..7th.ed..2010.2015.         0.874233      1.144 0.78242 0.976815
Derived.AJCC.M..7th.ed..2010.2015.         0.471214      2.122 0.29674 0.748284
Regional.nodes.positive..1988..            1.001136      0.999 0.99941 1.002859
white                                      1.430299      0.699 1.13945 1.795384
indian                                     0.833652      1.200 0.30487 2.279582
asian                                      0.671820      1.488 0.46635 0.967819
married                                    0.588520      1.699 0.50683 0.683384
separated                                  0.813847      1.229 0.41924 1.579872
unmarried                                  0.455192      2.197 0.06348 3.264157
widowed                                    0.753113      1.328 0.62938 0.901178
```

The results of z-test and 95% confidence intervals are shown in the figure respectively. The figure shows that age, gender, and SEER.Com bined. Mets. Palawan DX. Brian.. 2010.. , SEER.Com bined. Mets. Ats. DX. Lung.. 2010.. , Derived. AJCC. Stage. The Group.. 7th.ed.. 2010.2015., Derived. AJCC. T.. 7th.ed.. 2010.2015., Regional. Nodes. Positive.. 1988.. White these factors are risk factors, and their exponential regression coefficients are all greater than 1, indicating that the larger these variables are, the greater the risk of death will be. Other variables are not risk factors, and their exponential regression coefficients are all less than 1, indicating that the larger these variables are, the lower the risk of death is. If the exponential regression element of age is 1.059, it means that for every 1 increase in age, the risk of death is 1.059 times of the original. As a Grade.. The exponential regression coefficient of thru.2017 is 0.983, indicating that Grade.. Thru.2017. For every 1 increase, the risk of death increases by 0.983 times.

## 3.5 A comparison of whether to use principal component analysis

It can be seen that the competitive risk model without principal component can more intuitively see the impact of each factor on the death risk of the dependent variable compared with the competitive risk model with principal component. And because for different causes of death, different factors may present different results. Such as Grade.. Thru.2017. It is a risk factor in Colon exclusive Rectum, Rectum and Rectosigmoid Junction, Other Cause of Death, while in Diseases of Heart, Chronic Obstructive Pulmonary Disease and Allied Cond is not a risk factor. However, when there are too many data variables, it is easy to lead to insufficient accuracy of computer calculation, and it is easy to produce errors and fail to run, so that all variables cannot be included in the middle of the model.

```
> f1 <- FGR(Hist(Survival.months,COD)~.,data=datayuan,cause=7)
Error in solve.default(h, z[[2]]) :
  Lapack routine dgesv: system is exactly singular: U[37,37] = 0

> f1 <- FGR(Hist(Survival.months,COD)~Age.recode.with.single.ages.and.100.+Sex+Grade..thru.2017.
+         +Chemotherapy.recode..yes..no.unk.+SEER.Combined.Mets.at.DX.bone..2010..+
+           SEER.Combined.Mets.at.DX.brain..2010..+SEER.Combined.Mets.at.DX.liver..2010..+
+           SEER.Combined.Mets.at.DX.lung..2010..+Derived.AJCC.Stage.Group..7th.ed..2010.2015.+
+           Derived.AJCC.T..7th.ed..2010.2015.+Derived.AJCC.N..7th.ed..2010.2015.+
+           Derived.AJCC.M..7th.ed..2010.2015.+Radioactive_implants+
+           Radiation_NOS_method_or_source_not_specified+Radioisotopes+Regional.nodes.positive..1988..
+         +white+indian+asian+distant+Regional+Localized+Unstaged+
+           married+separated+unmarried+widowed+rx10_19+rx20_80+rx90+rx99,data=datayuan,cause=7)
Error in solve.default(h, z[[2]]) :
  system is computationally singular: reciprocal condition number = 1.49281e-19
```

As shown in the figure above, if we use the survival month and COD to directly make a competitive risk model for other variables in the data set, an error will be reported and it cannot be solved. The main reason is that irreversible matrices or calculation problems occur in the calculation process.

It can be seen that the competitive risk model without principal component can more intuitively see the impact of each factor on the death risk of the dependent variable compared with the competitive risk model with principal component. However, when there are too many data variables, it is easy to lead to insufficient accuracy of computer calculation, and it is easy to produce errors and fail to run, so that all variables cannot be included in the middle of the model. This makes it impossible to directly compare the results of pca without pca with those using PCA. In general, using principal component analysis can reduce the dimension of data. We can observe the effect of each variable on the risk of death by taking the results of the principal component further to look at the main components. However, the model without principal components are prone to errors and cannot be run. The number of variables can only be appropriately reduced, and the impact of all variables on death risk cannot be observed at the same time.

## 3.6 Prediction based on Fine Gray competitive risk model

Based on the results of the Fine Gray competitive risk model, the levels of 1-year, 3-year, and 5-year survival were predicted for patients with different influencing factors.

For example, for patient A, the RC values of his influencing factors are: 2,2,1,0.5,1,5,0.1,2,6,5 The cumulative probability of patient A's death due to Colon exclusive Rectum is shown in Figure 3.

Figure 3 Death probability of Patient A

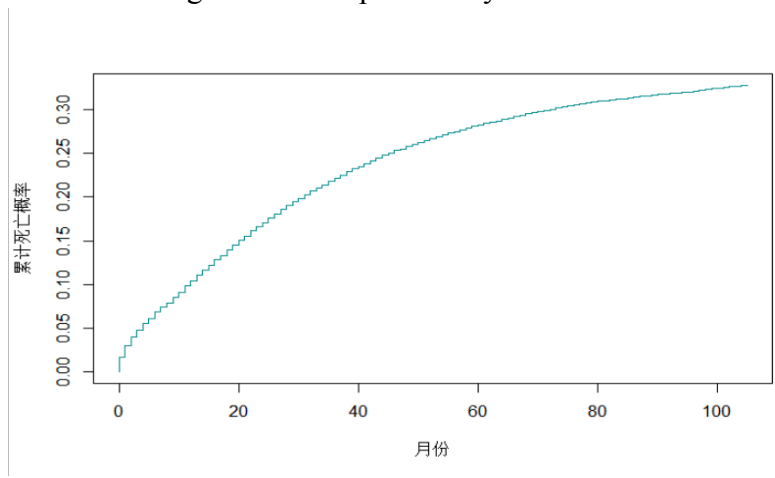

## 3.7 Conclusion

The competitive risk model is successfully constructed, which can predict different diseases.

The innovation of this study is as follows: considering that the existence of competitive risk in patients may affect the survival of patients and the analysis of influencing factors, the fine-Gray competitive risk model is innovatively used to analyze, avoiding the result deviation caused by the traditional survival analysis method. The large amount of patient survival information provided in the database was used to establish the model, which made the model more stable. On the question of whether to use principal component analysis, it is best to analyze the problem in two ways and draw some different constructive conclusions. If only one of these methods is adopted, the conclusion drawn will be **one-sided**.

# 4.Multiple Linear regression analysis

## 4.1 Model introduction

Multiple Linear regression analysis is a quantitative depiction of linear dependencies between one dependent variable and multiple independent variables. The basic idea of regression analysis is that although there is no strict, deterministic functional relationship between independent variables and dependent variables, one can try to find a mathematical expression that best represents the relationship between them. Specifically, multiple linear regression analysis mainly solves the following problems.

(1) Determine whether there is a correlation between several specific variables, and if so, find out the appropriate mathematical expressions between them;
(2) According to the value of one or several variables, predict or control the value of another variable, and can know what kind of accuracy this prediction or control can achieve;

The survival time was predicted by multiple linear regression. The specific model is as follows:
$$Y = X\beta + \varepsilon$$

Where Y is the survival time, Independent variable X is a matrix composed of 10 components from RC1 to RC10 of the previous principal component analysis. β is the regression coefficient vector, and the size of β indicates the influence of each independent variable on the dependent variable, and ε is the residual.

When establishing pluralistic regression model, in order to ensure that the regression model has excellent explanatory ability and prediction effect, we should first pay attention to the selection of independent variables, and the criteria are as follows:
(1) The independent variable must have a significant influence on the dependent variable and show a close linear correlation;
(2) The linear correlation between independent variables and dependent variables must be real, rather than formal;
(3) There should be mutual exclusion between independent variables, that is, the degree of correlation between independent variables should not be higher than the degree of correlation between independent variables and the cause of dependent variables;
(4) Independent variables should have complete statistical data, and their predicted values are easy to determine.

## 4.2 Results of the correlation

Firstly, correlation analysis was conducted between dependent variables and independent variables, as shown in Figure 4:

Figure 4 Correlation between survival time and various variables

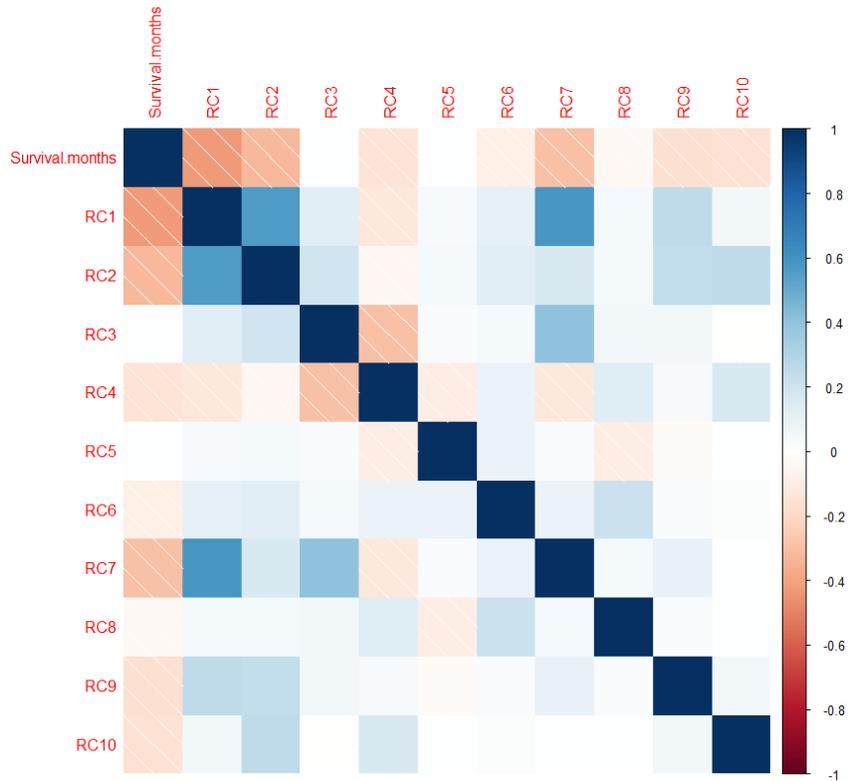

It can be seen from Figure 1 that survival time is correlated with most variables. Among them, RC1,RC2 and RC7 have the highest degree of correlation with negative correlation, while RC3,RC5 and RC8 have low correlation (close to 0).

Therefore, the correlation test between the survival time and RC3,RC5 and RC8 is conducted in this paper to determine whether their correlation is 0. If so, it is considered that they can be excluded from the model.

```
> cor.test(data5$Survival.months,data5$RC3)

        Pearson's product-moment correlation

data:  data5$Survival.months and data5$RC3
t = 3.2549, df = 119236, p-value = 0.001135
alternative hypothesis: true correlation is not equal to 0
95 percent confidence interval:
 0.003749979 0.015100949
sample estimates:
        cor
0.009425768
```

According to the correlation test, there was a significant positive correlation between survival time and RC3 (P =0.0011 < 0.05).

```
> cor.test(data5$Survival.months,data5$RC5)

        Pearson's product-moment correlation

data:  data5$Survival.months and data5$RC5
t = 2.9522, df = 119236, p-value = 0.003156
alternative hypothesis: true correlation is not equal to 0
95 percent confidence interval:
 0.002873383 0.014224531
sample estimates:
        cor
0.008549232
```

According to the correlation test, there was a significant positive correlation between survival time and RC5 (P =0.0031 < 0.05).

```
> cor.test(data5$Survival.months,data5$RC8)

        Pearson's product-moment correlation

data:  data5$Survival.months and data5$RC8
t = -11.527, df = 119236, p-value < 2.2e-16
alternative hypothesis: true correlation is not equal to 0
95 percent confidence interval:
 -0.03903172 -0.02769238
sample estimates:
        cor
-0.03336313
```

According to the correlation test, there was a significant positive correlation between survival time and RC8 (P < 0.05).

In order to facilitate the subsequent test and prediction of the model, we divided the data into three parts. The first part is the surviving data. There are 69,476 samples in this part, which we will use for prediction, called the prediction set. Then, the remaining samples are divided into two parts by random number random sampling method. The second part accounts for 80% of the remaining data, with a total of 39,809 samples. This data set will be used to build our regression model, which we call training set. The third part accounts for 20% of the data and contains 9953 samples. This data set will be used to test the validity of the model, which we call the validation set.

## 4.3 Regression analysis results

Using the training set for regression analysis, the results are as follows:

```
> summary(ols)

Call:
lm(formula = Survival.months ~ RC1 + RC2 + RC3 + RC4 + RC5 +
    RC6 + RC7 + RC8 + RC9 + RC10, data = subdata)

Residuals:
    Min      1Q  Median      3Q     Max
-469.16  -71.85    2.32   75.78  356.41

Coefficients:
             Estimate Std. Error  t value Pr(>|t|)
(Intercept) 845.3313     0.6517 1297.115   <2e-16 ***
RC1          -3.0746     0.1530  -20.101   <2e-16 ***
RC2          -0.1246     0.1530   -0.814    0.416
RC3          -4.8133     0.2811  -17.120   <2e-16 ***
RC4          52.0712     0.2998  173.699   <2e-16 ***
RC5          -4.0100     0.3590  -11.171   <2e-16 ***
RC6         -28.5801     0.3634  -78.645   <2e-16 ***
RC7          -3.5023     0.2375  -14.749   <2e-16 ***
RC8         -12.0188     0.4551  -26.412   <2e-16 ***
RC9           6.8865     0.3626   18.993   <2e-16 ***
RC10         12.1525     0.4386   27.707   <2e-16 ***
---
Signif. codes:  0 '***' 0.001 '**' 0.01 '*' 0.05 '.' 0.1 ' ' 1

Residual standard error: 109.3 on 39798 degrees of freedom
Multiple R-squared:  0.6078,	Adjusted R-squared:  0.6077
F-statistic:  6168 on 10 and 39798 DF,  p-value: < 2.2e-16
```

According to the results of multiple regression, the regression coefficients of most variables were significant at the confidence level of 0.05. This indicates that the regression coefficients of most independent variables are not 0, that is, they all have an impact on the dependent variable. When other variables remain unchanged, the survival time will be reduced by 3.07 months on average for every 1 unit increase of RC1. Holding other variables constant, the survival time decreased by 0.12 months for each 1 unit increase in RC2. Holding other variables constant, a 1-unit increase in RC3 would reduce the survival time by 4.81 months on average. With other variables unchanged, survival time increased by 52.07 months on average for every 1 unit increase in RC4. Holding other variables constant, an increase of RC5 by 1 unit will reduce the survival time by 4.01 months on average. Holding other variables constant, the survival time was reduced by 28.58 months on average for each 1 unit increase in RC6. Holding all other variables constant, an increase of 1 unit in RC7 resulted in an average reduction of 3.50 months in survival. Holding other variables constant, survival time was reduced by 12.01 months for each 1 unit increase in RC8. Holding other variables constant, a 1-unit increase in RC9 would increase the survival time by 6.89 months on average. Holding all other variables constant, a 1-unit increase in RC10 resulted in an average increase in survival time of 12.15 months. The overall F-test of the model was 6168 (p value < 0.05), indicating that the regression coefficients of model variables were 0 at different times. The R square of the model is 0.61, and the adjusted R square is 0.61, indicating that independent variables explain 61% of the variation of the dependent variable. This shows that there is no problem in the model setting and the model fitting effect is very good. In the follow-up, the regression model was tested as shown in the figure below.

Figure 5 Model test diagram

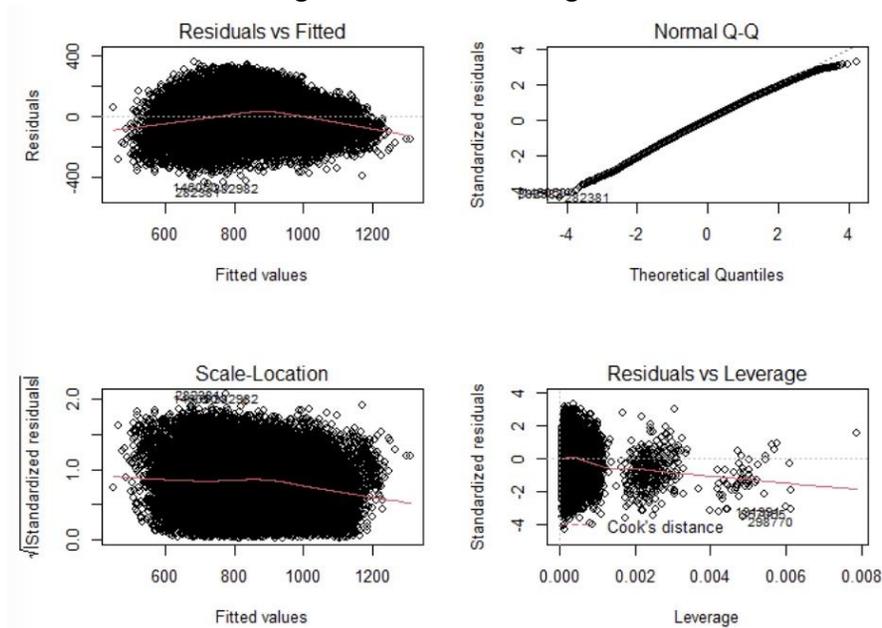

Check the model as shown in Figure 5. A straight line in the upper left corner shows that the model conforms to the linear hypothesis and does not need to add higher-order terms. In the upper right corner is the residual QQ map, most of the points in the figure fall on diagonal dotted lines, indicating that the residual conforms to the assumption of normality. The scattered points in the lower left corner are relatively scattered, indicating that the model conforms to the assumption of homovariance. As can be seen from the figure at the lower right corner, there are no outliers in the model data.

Further, the multicollinearity test of the model is shown as follows.

```
> vif(ols)
     RC1      RC2      RC3      RC4      RC5      RC6      RC7      RC8      RC9     RC10
2.953406 2.054526 1.396484 1.291559 1.043317 1.080350 2.000055 1.063710 1.127840 1.192170
```

The test results show that the variance inflation factor (VIF) of each variable is less than 10, which indicates that there is no linear relationship between each independent variable, there is no multicollinearity in the model, and the model fits well. The final model is:

$$Survival.months = 845.33 - 3.07RC1 - 0.12RC2 - 4.81RC3 + 52.07RC4 - 4.01RC5 - 28.58RC6 - 3.50RC7 - 12.02RC8 + 6.89RC9 + 12.15RC10$$

## 4.4 Multicollinearity judgment

### 4.4.1 Definition

If the t test of a certain regression coefficient fails, it may be caused by the insignificant shadow level of the independent variable corresponding to the coefficient on the dependent variable. In this case, the independent variable should be removed from the regression model, and a simpler regression model should be established again or the independent variable should be replaced. It may also be caused by collinearity between independent variables. At this time, we should try to reduce the influence of collinearity.

Multicollinearity means that in multiple linear regression equations, there is a strong linear relationship between independent variables. If this relationship exceeds the linear relationship between dependent variables and independent variables, the stability of the regression model will be damaged and the estimation of regression coefficients will be inaccurate. It should be pointed out that multicollinearity is unavoidable in multiple regression models, as long as the multicollinearity is not too severe. To determine whether there is severe multicollinearity in multiple linear regression equations, the determinable coefficient R between each two independent variables can be calculated separately. If **R2 >R1** or close to R2, the influence of multiple linearity should be reduced. The condition number k= $\lambda_1/\lambda_p$ ($\lambda_1$ is the maximum eigenvalue, $\lambda_p$ is the minimum eigenvalue) of the correlation coefficient matrix between independent variables can also be calculated, and k<100, then there is no multi-point collinearity. If 100≤k≤1000, there is strong multicollinearity between the independent variables; if k>1000, there is serious multicollinearity between the independent variables. The main way to reduce multicollinearity is to change the value of the independent variable, such as changing the absolute number to relative number or average number, or replacing other independent variables.

### 4.4.2 Example

In addition, this paper also carries on multiple linear regression analysis to the original data, using 42 original data as independent variables to do multiple linear regression to the dependent variable survival time.

Before regression, race variables, marriage status variables and other multi-factor variables should be deleted to avoid multicollinearity (i.e., there is a linear relationship between x independent variables, and the sum of several race variables is 1, which will affect the regression result).

$$x_1 = 2x_2$$

If the multicollinearity problem is not avoided, the interpretation of the model results will be greatly biased. For example, if there is a linear relationship between two variables, then all these variables are included in the model. These variables satisfy the following formula.

$$y = x_1 + x_2$$

Then there will be countless solutions for the regression coefficients of these variables. For example, we assume that the regression coefficients of these four variables are 1.
Then the result of the calculation may be (not only the following two, but any result that satisfies the above equation may be possible)

$$y = 1.5x_1 + 0.0x_2 \; ; \; y = 0.5x_1 + 2.0x_2$$

This will make our estimation of the regression coefficient have a big bias, which will lead to a big deviation between our estimation and the actual result.

### 4.4.3 Treatment method in this paper

Multicollinearity is not a unique phenomenon of multiple linear regression, but may also exist in multiple Logistic regression and Cox regression. In the final analysis, both Logistic regression and Cox regression can be classified as Generalized Linear models, while models containing two or more independent variables can be called multi-factor models. Therefore, multi-factor Generalized linear models may face multicollinearity problems.

Multicollinearity is a phenomenon that may exist between independent variables and may exist objectively. The reason why we discuss it is called a problem, because collinearity will bring great uncertainty to our parameter estimation. In medical research, the effect relationship between exposure and outcome is generally explored, so the attention to regression coefficient is much higher than other model indicators (such as R2, etc.). With an understanding of multicollinearity and its hazards, it is important to be aware of this phenomenon when conducting multifactorial data analysis.

And for this kind of multicollinearity, there are some methods to solve multicollinearity
**(1) Expand the sample size**
Multicollinearity is essentially a problem of data. Theoretically highly correlated variables may not have highly correlated observed values, and vice versa. Therefore, it is possible to eliminate or mitigate the multicollinearity problem by enlarging the sample size, increasing observations, using different data sets, or adopting new samples.
**(2) Impose some constraints on the model**
In the model with multicollinearity, the variance of the coefficient estimator can be reduced by imposing some constraints according to economic theory, such as adding the constraint of constant economies of scale into the Cobb-Douglas production function, which can solve the multicollinearity problem caused by the high correlation between capital and labor.
**(3) Delete one or more collinear variables**
In this way, we can actually estimate fewer parameters with the given data, thus reducing the need for observation information to solve the multicollinearity problem. Which variables to delete can be determined based on the results of hypothesis testing. However, it should be noted that this approach will lead to the deviation of the estimation results and cause the problem of missing variables, so it should be used with caution.

**(4) Properly deform the model**

In the model, two highly correlated variables can be mathematically deformed, such as dividing two highly correlated variables to solve the multicollinearity problem caused by these two variables.

**(5) Principal component regression**

Practice is to explain all variables using principal component analysis for principal components, each of all the main component is a linear combination of the explanatory variables, because each principal component between unrelated, and can use a few principal components can explain most of the variance of X variables, thus the multicollinearity, can replace the original variable with principal component regression calculation.

**Principles for dealing with multicollinearity problems**

Multicollinearity is universal, and minor multicollinearity problems can be left unchecked. Severe multicollinearity problems can generally be found empirically or by regression results. Important explanatory variables, such as influence symbols, have low values. Necessary measures should be taken according to different circumstances.

For our model, there is enough data in the first place that method 1 does not apply. **Method 2** and **Method 4** are both based on the specific form of the model (for example, the relationship between height and weight has been determined, and the size of the relationship is determined by the model). However, there is no established model in this paper. Therefore, **Method 2** and **Method 4** are not applicable. If the principal component regression is not adopted, the elimination of variables will be the most direct and effective method.

## 4.5 When not using principal component analysis

The final regression results are shown below.

By comparing the results of principal component analysis with and without principal component analysis, we found that the directions of principal component analysis and results without principal component were mostly the same, but there were still some differences. We analyze the coefficient difference between the main components of each principal component in the two regression methods. (main component refers to the principal component in the rotation matrix, the weight of the variable to a principal components factor is greater than or equal to 0.6)

The main components of RC1 are SEER.Com bined. Mets. Palawan DX. Liver.. 2010.. , Derived. AJCC. M.. 7th.ed.. 2010.2015.,, their coefficients in multiple linear regression are -1.59, 0.87 and -0.84 respectively. After the principal component is used, the regression coefficient of RC1 is -3.07, reflecting that they are negatively correlated with the dependent variable.

The main composition of RC2 is Regional. Nodes. Positive.. 1988.. Rx0, rx20_80. When they join the model alone, Regional. Nodes. Positive.. 1988.. The coefficient of is -0.05, rx0 is deleted to avoid multicollinearity, and the regression coefficient of RX20-80 is 7.20. After the principal component was used, the regression coefficient of RC2 was -0.12, indicating that other components in RC2 offset the positive correlation of RX20_80, thus making RC2 negative.

The main components of RC3 are Chemotherapy.recode..yes..No.Unk.,beam_radiation,Recommended_unknown_if_administered, and their regression coefficients are 9.28, -1.16 (beam_radiation is deleted to avoid collinearity), The regression coefficient of RC3 is -4.81, reflecting the negative correlation between the main components of these variables and the dependent variable.

The main components of RC4 are the Age. The recode. With the single. Which. And. 100., widowed. Their regression coefficient is 11.83 and 0.08, the dependent variable is a positive correlation; The regression coefficient of RC4 is 52.07, which is also positive correlation.

The main components of RC5 are White and Asian, which have opposite effects on RC5. When constituting the principal component, White is negatively correlated with RC5, while Asian is positively correlated with RC5. In the regression, the coefficients of both of them are positive, 1.06 and 2.85 respectively. The regression coefficient of RC5 is -4.01, indicating that RC5 still reflects the negative correlation between their main components and the dependent variable.

The main components of RC6 are married and single, among which married has a negative correlation with RC6 and single has a positive correlation with RC6. In the linear regression, the regression coefficients of the two variables are 2.18 and -0.69, and the regression coefficient of RC6 is -28.58, which well reflects the relationship between these two variables and the dependent variable.

The main components of RC7 is Derived. The AJCC. Stage. The Group.. 7th.ed.. 2010.2015., Derived. AJCC. T.. 7th.ed.. 2010.2015., Derived. AJCC. N.. 7th.ed.. There are 5 variables in total, of which only remuneration is negatively correlated with RC7, while the others are positively correlated. Their regression coefficients were -1.74, -1.11, -1.85, 5.08 and 3.16, respectively. The regression coefficient of RC7 was -3.50, indicating that the changes of other variables offset the positive correlation between monetary and Regional, and RC7 showed a significant negative correlation.

The main components of RC8 are divorced, married. The regression coefficient of Married is 2.18, divorced is deleted because of avoiding multicollinearity, and the regression coefficient of RC8 is -12.02, reflecting that the common effect of the two variables on the dependent variable is negatively correlated.

The main components of RC9 are SEER.Com bined. Mets. Palawan DX. Ipads.. 2010.. , SEER.Com bined. Mets. Ats. DX. Brian.. 2010.. , their regression coefficients for the dependent variable are -3.14 and -5.18, and that for RC9 is 6.88, reflecting their positive correlation with the dependent variable.

The main components of RC10 are Unstaged and RX99, and their regression coefficients are positive, while RC10's regression coefficients are negative, indicating that other components of RC10 offset the positive correlation of RX99.

```
Coefficients:
                                                   Estimate Std. Error   t value Pr(>|t|)
(Intercept)                                       45.705804   1.359709    33.614  < 2e-16 ***
Age.recode.with.single.ages.and.100.              11.831195   0.008702  1359.660  < 2e-16 ***
Sex                                               -0.987033   0.211916    -4.658 3.21e-06 ***
Grade..thru.2017.                                 -4.071397   0.161653   -25.186  < 2e-16 ***
Chemotherapy.recode..yes..no.unk.                  9.285832   0.258995    35.853  < 2e-16 ***
SEER.Combined.Mets.at.DX.bone..2010..             -3.141080   0.743579    -4.224 2.40e-05 ***
SEER.Combined.Mets.at.DX.brain..2010..            -5.182535   1.457825    -3.555 0.000378 ***
SEER.Combined.Mets.at.DX.liver..2010..            -1.591122   0.375191    -4.241 2.23e-05 ***
SEER.Combined.Mets.at.DX.lung..2010..             -0.093921   0.397164    -0.236 0.813062
Derived.AJCC.Stage.Group..7th.ed..2010.2015.      -1.736286   0.283321    -6.128 8.96e-10 ***
Derived.AJCC.T..7th.ed..2010.2015.                -1.106595   0.115714    -9.563  < 2e-16 ***
Derived.AJCC.N..7th.ed..2010.2015.                -1.843948   0.137741   -13.387  < 2e-16 ***
Derived.AJCC.M..7th.ed..2010.2015.                 0.878039   0.968199     0.907 0.364476
Regional.nodes.positive..1988...1                 -0.047533   0.004881    -9.738  < 2e-16 ***
CS.mets.at.dx..2004.2015.                         -0.076878   0.012690    -6.058 1.39e-09 ***
white                                              1.062722   0.304194     3.494 0.000477 ***
asian                                              2.854148   0.457620     6.237 4.51e-10 ***
indian                                            -1.876235   1.193354    -1.572 0.115904
married                                            2.175710   0.343110     6.341 2.30e-10 ***
separated                                          1.006028   0.971743     1.035 0.300544
single                                            -0.696149   0.388542    -1.792 0.073189 .
unmarried                                         -2.048133   2.281703    -0.898 0.369386
widowed                                            0.079330   0.400792     0.198 0.843099
Combination_of_beam_with_implants_or_isotopes     -1.382714   7.548740    -0.183 0.854664
Radiation_NOS_method_or_source_not_specified       0.989273   2.024440     0.489 0.625081
Radioactive_implants                               5.306585   3.388848     1.566 0.117381
Radioisotopes                                     -0.810590   3.491369    -0.232 0.816408
Recommended_unknown_if_administered               -1.162579   0.327074    -3.554 0.000379 ***
Refused                                            0.131586   1.314301     0.100 0.920251
distant                                           -0.842648   1.256602    -0.671 0.502494
Localized                                          5.088181   0.953623     5.336 9.57e-08 ***
Regional                                           3.164598   1.107030     2.859 0.004257 **
rx10_19                                            4.544332   7.064275     0.643 0.520044
rx20_80                                            7.201086   0.500022    14.402  < 2e-16 ***
rx90                                               7.040607   2.150388     3.274 0.001061 **
rx99                                               2.378376   2.679007     0.888 0.374663
---
Signif. codes:  0 '***' 0.001 '**' 0.01 '*' 0.05 '.' 0.1 ' ' 1

Residual standard error: 19.95 on 39773 degrees of freedom
Multiple R-squared:  0.9869,    Adjusted R-squared:  0.9869
F-statistic: 8.588e+04 on 35 and 39773 DF,  p-value: < 2.2e-16
```

To sum up, we believe that the principal component reflects the change of most independent variables to dependent variables, and there are still a small part of independent variables that are contrary to the result of the principal component, which is offset by other components in the principal component. Then we get the model fitting results without principal component analysis as follows.

```
Residual standard error: 19.95 on 39773 degrees of freedom
Multiple R-squared:  0.9869,	Adjusted R-squared:  0.9869
F-statistic: 8.588e+04 on 35 and 39773 DF,  p-value: < 2.2e-16
```

From the model fitting results, the results of using principal component analysis and not using principal component analysis were compared. When pca was not used, the R2 of the model was higher and the square of R reached 0.98, indicating that the use of raw data could explain the variation of dependent variable survival time more than that of PCA. However, in terms of the use of comparative variables, only 10 independent variables are used after principal component analysis, while there are many variables in the original data. The use of principal components well reflects the correlation between most factors and dependent variables. This comparison is better after using principal component analysis.

# 5. Testing

## 5.1 Testing result

We first used the test set to predict the model using principal component regression, that is, a total of 10 variables rc1-RC10 were used to predict the survival time. Our judgment criterion is that if the confidence interval of the predicted results of the model includes the actual results of the data, the prediction of the model is considered to be effective for this sample. Finally, the proportion of the sample with effective prediction is calculated to illustrate the prediction effect of the model.

Confidence interval is a method of interval estimation in statistics. [a, b] is used to represent the error range of the sample's estimation of the population mean. Since the exact values of a and B depend on the degree of confidence we want to have in the result that the interval contains the population mean, this interval is called the confidence interval.

To put it simply, when we estimate a variable, due to the influence of random error, our point estimate value often has a certain difference with the actual value, which may be large or small, or positive or negative. However, a variable usually has a probability distribution, for example, the probability that the data conforming to the normal distribution will be within ±2 standard deviations of its mean. So we can make an interval estimate of the variable we want based on the data, which means that we estimate that this point has a high probability of falling within this interval. For example, if we estimate the 95% confidence interval of a number to be [-1,1], we assume that the number is 95% likely to be within this interval. If the interval we estimate doesn't contain the actual numbers, that's also a small probability. But if this happens in a very large proportion, then our model doesn't fit. In addition, when estimating the interval of data, it will also be affected by the data itself, because if the distribution range of a data is very wide and the variance is very large, it indicates that the possible range of the data is also very large. In this case, the range of the estimated confidence interval is also very large. Conversely, if the data is very concentrated and the variance is very small, then the estimated confidence interval is also very small.

In addition, we also calculate the mean square error of the prediction model. The mean square error is the sum of the squares of our actual time to live minus our predicted time to live, as shown below. This data can explain the deviation degree between the prediction result of our model and the actual result. The smaller this number is, the better the prediction ability of our model is. Of course, there is no unified standard to show that the prediction effect of the model is better when the number reaches many hours, because there are also the influence of random error and dispersion degree (variance) of existing data. Just like fitting a straight line to a scatter graph, you can only try to distribute most points on both sides of the line, and there is no way to make all points fall on the line.

$$MSE = \sum (y - \hat{y})^2$$

Where MSE represents the mean square error, y represents the actual dependent variable and represents the dependent variable predicted by the model.

We used the test set pair to predict the principal component regression model and the regression model without principal component respectively (since the test set also has 23848 samples, only the first 10 lines of the predicted results are shown here). The following are the predicted results of the model with principal component regression.

```
> fore1
   Point Forecast    Lo 80      Hi 80     Lo 95      Hi 95
1       1033.7962  893.6951  1173.8973  819.5271  1248.0653
2       1041.3674  901.2623  1181.4724  827.0922  1255.6425
3        750.4811  610.3772   890.5850  536.2078   964.7545
4        972.8248  832.6823  1112.9673  758.4923  1187.1572
5        960.1941  820.0941  1100.2940  745.9267  1174.4614
6        837.8320  697.7256   977.9383  623.5549  1052.1091
7        781.3188  641.1878   921.4498  567.0040   995.6336
8        826.7571  686.6360   966.8783  612.4573  1041.0569
9        813.7447  673.6391   953.8504  599.4686  1028.0208
10      1090.2295  950.1234  1230.3355  875.9527  1304.5062
```

As can be seen from the figure, the prediction results are divided into 5 columns. The first column is the Point Forecast, which is the result of Point estimation, that is, the estimated value obtained after using the principal component regression model for prediction. The second and third columns are Lo 80 and Hi 80, respectively. These are the lower and upper limits of the 80% confidence interval, respectively, indicating that 80% of the predicted dependent variable will fall within this interval. The last two columns are Lo 95 and Hi 95. These are the lower and upper limits of the 95% confidence interval, respectively, indicating a 95% probability that the predicted dependent variable will fall within this interval. The 95% confidence interval is definitely larger and broader than the 80% confidence interval. To analyze the predictive power of the model, we calculated the proportion that included the actual survival time in the 80% confidence interval and the proportion that included the actual survival time in the 95% confidence interval.

Finally, we obtained that in the prediction of the inclusive principal component model, 80% confidence interval contained 79.90% survival time, 95% confidence interval contained 95.42% survival time. It shows that the prediction effect of the model is good and the range of survival time can be predicted. The predicted mean square error of the first model is 16506587, which seems to be very large. However, since there are nearly 10,000 samples of data and the mean square error is the sum of the squares of the errors, this value is actually not very large.

We further forecast the regression model without principal component, and the results are as follows.

```
> fore2
   Point Forecast     Lo 80     Hi 80      Lo 95     Hi 95
1         961.8554  936.2771   987.4336  922.7363 1000.9745
2        1011.7713  986.1450  1037.3975  972.5787 1050.9638
3         603.9019  578.3258   629.4779  564.7862  643.0175
4         997.3495  971.7658  1022.9332  958.2220 1036.4769
5        1013.3702  987.7934  1038.9469  974.2534 1052.4870
6         997.1703  971.5912  1022.7495  958.0498 1036.2908
7         893.1920  867.6093   918.7746  854.0662  932.3178
8         972.6771  947.0936   998.2607  933.5499 1011.8044
9         759.2191  733.6415   784.7967  720.1011  798.3372
10       1090.4644 1064.8889  1116.0399 1051.3495 1129.5793
```

As you can see, the numbers alone do not determine which of the two models has the better predictive power. Therefore, we also use the same method to analyze the prediction effect of this model.

Finally, we obtained that in the prediction without the principal component model, 80% confidence interval contained 82.23% survival time, and 95% confidence interval contained 94.67% survival time. It shows that the prediction effect of the model is good and the range of survival time can be predicted. The predicted mean square error of the second model is 15809981, which seems to be very large. However, as the data volume is close to 10,000 and the mean square error is the sum of the squares of the errors, this value is actually not very large.

By comparing the prediction ability of the two models, we find that the model using principal components is slightly better than the model without principal components in interval prediction. In terms of the deviation from the actual value, using a model without principal components will be relatively more accurate. Of course, considering that the model without principal component uses more independent variables, while the model with principal component only has 10 independent variables, we believe that the model with principal component is better than the model without principal component.

## 5.2 Conclusion

In this paper, the correlation between the survival time and each principal component is analyzed, and the regression model of survival time is established by multiple linear regression. Except RC3,RC5 and RC8, the other variables can shorten the survival time.
In this paper, the differences between the two models in model fitting results and survival time prediction were compared. Finally, it is concluded that the use of principal component analysis can greatly reduce the dimension of data, but the overall impact on the model is relatively small, and the use of principal component analysis after regression model is a better method.

# 6. Predictive missing value

After the above tests, we believe that our model has good predictive ability and can be used to predict data. We analyze the prediction set with a model that includes principal component analysis. The prediction results of the model are as follows:

```
> forecast1
   Point Forecast      Lo 80      Hi 80      Lo 95      Hi 95
1        850.0831   709.9677   990.1985   635.7921  1064.3741
2        883.6833   743.5752  1023.7914   669.4035  1097.9631
3        972.1230   831.9884  1112.2577   757.8026  1186.4435
4        929.3010   789.1853  1069.4166   715.0096  1143.5923
5        764.6926   624.5659   904.8193   550.3843   979.0008
6        757.5560   617.4528   897.6591   543.2837   971.8282
7        784.7229   644.6139   924.8319   570.4417   999.0040
8        879.9078   739.7881  1020.0275   665.6103  1094.2053
9        864.9659   724.8595  1005.0723   650.6887  1079.2431
10       685.1694   545.0460   825.2929   470.8662   899.4727
```

Finally, we used the inclusion principal component model to predict the prediction set, and obtained the predicted average survival time and 80% and 95% confidence interval.
We have made a histogram of the data and forecast results as shown below.

Figure 6 The data and forecast results

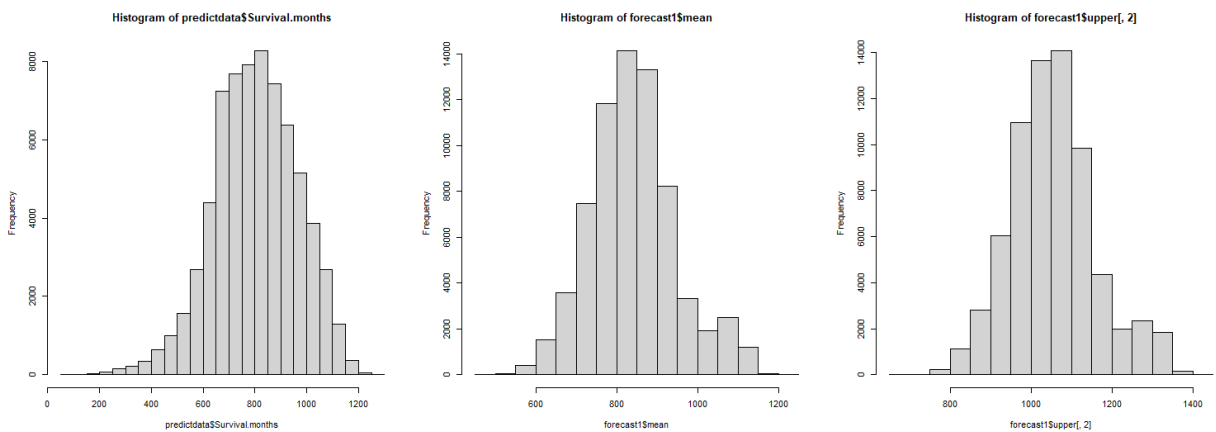

The graph on the left shows the survival time of the original survival data, the graph in the middle shows the survival time predicted by the model including principal component analysis, and the graph on the right shows the maximum 95% confidence interval predicted by the model including principal component analysis. As can be seen from the figure, the predicted survival time is concentrated at about 850 months, which is quite consistent with the original data. The predicted confidence interval is mostly within 1100 months.

We analyze the prediction set with a model that does not include principal component analysis. The prediction results of the model are as follows:

```
> forecast2
   Point Forecast     Lo 80     Hi 80     Lo 95     Hi 95
1        827.4547  801.8759  853.0335  788.3348  866.5747
2        991.6024  966.0251 1017.1797  952.4848 1030.7200
3        997.3301  971.7478 1022.9123  958.2049 1036.4553
4        932.3432  906.7629  957.9235  893.2210  971.4655
5        665.9398  640.3589  691.5207  626.8167  705.0629
6        796.5767  771.0011  822.1523  757.4617  835.6918
7        784.6429  759.0636  810.2221  745.5223  823.7635
8        884.7551  859.1749  910.3352  845.6331  923.8770
9        733.2009  707.6248  758.7771  694.0851  772.3168
10       669.2965  643.7123  694.8808  630.1683  708.4248
```

Finally, we used the prediction model without principal component to predict the prediction set, and obtained the predicted average survival time and 80% and 95% confidence interval.
We have made a histogram of the data and forecast results as shown below.

Figure 7 The data and forecast results

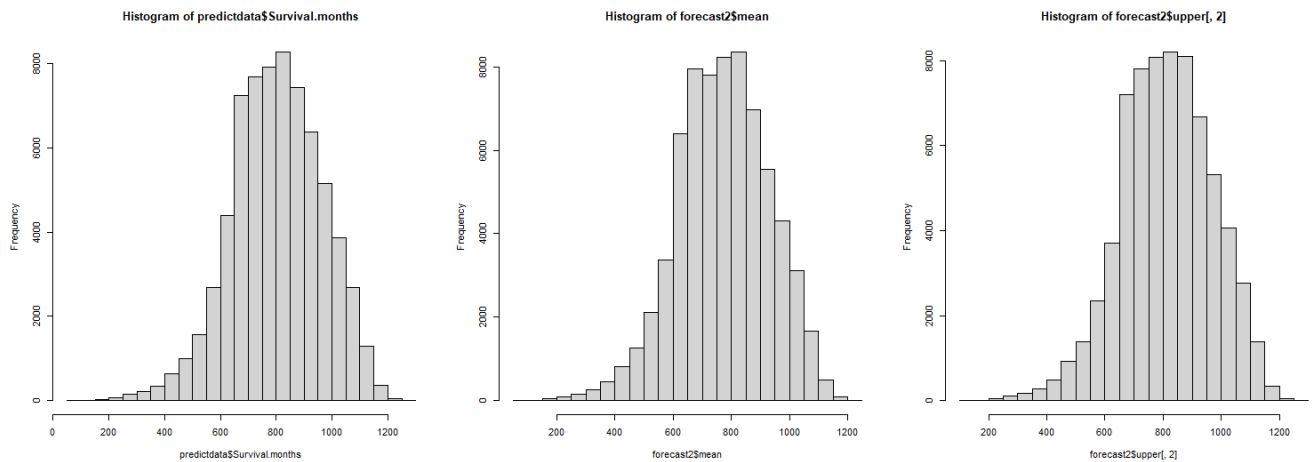

The graph on the left shows the survival time of the original survival data, the graph in the middle shows the survival time predicted by the model without principal component analysis, and the graph on the right shows the maximum 95% confidence interval predicted by the model without principal component analysis. As can be seen from the figure, the predicted survival time is concentrated around 800 months, which is quite consistent with the original data, and the predicted confidence interval is mostly within 900 months.

Comparing the two models, on the predictions of a survival time close to, 800 a month or so, and in the confidence interval range, do not include the model more pessimistic, principal component analysis, of course, the main reason is the use of more independent variables, some prediction accuracy is higher, so that the relatively small error range, forecast range is small. However, the model containing the principal component uses fewer variables, the prediction accuracy is fair, and the prediction range is larger.

# 7. Conclusion

(1) Principal component analysis has a significant effect. Principal component analysis can greatly reduce the data dimension, and more representative principal component factors can be added to the subsequent model.

(2) The competitive risk model has been successfully constructed, and the model has a good degree of differentiation and fit. This paper finds that for different diseases, each principal component factor will present different effects. Combining the results of several diseases with the largest number of cases, RC4 and RC10 had adverse effects on all five diseases, and increasing RC4 and RC10 led to an increased risk of death. RC1,RC3 and RC5 have a good effect on diseases outside Rectum and Rectosigmoid Junction, and individuals with higher RC1,RC3 and RC5 have a relatively low risk of death. RC2 had an adverse effect on diseases other than Colon exclusive Rectum, resulting in an increased risk of individual death. RC9 has adverse effects on diseases other than Rectum and Rectosigmoid Junction, resulting in an increased risk of individual death. However, RC6,RC7 and RC8 do not have significant effects on multiple diseases, so it can be considered that they have no impact on individual mortality risk.

(3) The linear regression model was successfully constructed. This paper found that RC3,RC5 and RC8 in the model were positively correlated with survival time, and were positive factors of survival time, while other principal component factors were negative factors of survival time, and were negatively correlated with survival time.

(4) The constructed competitive risk model can predict the survival risk of patients, while the constructed linear regression model can predict the survival time of patients.